\documentclass[11pt,a4paper]{article}

\usepackage[T1]{fontenc}
\usepackage[utf8]{inputenc}%
\usepackage[english]{babel}
\usepackage{lmodern}
\usepackage{microtype}

\usepackage[a4paper,margin=2.5cm]{geometry}
\usepackage{setspace}
\usepackage{amsmath,amssymb,mathtools}
\usepackage{graphicx}
\graphicspath{{figures/}}
\usepackage{booktabs}
\usepackage{caption}
\DeclareCaptionLabelSeparator{pipe}{\ $|$\ }
\usepackage{subcaption}
\usepackage{float}
\usepackage{braket}

\usepackage[super,sort&compress]{natbib}

\usepackage{authblk}

\usepackage{xcolor}
\usepackage[
  colorlinks=true,
  linkcolor=blue!50!black,
  citecolor=green!40!black,
  urlcolor=blue!60!black
]{hyperref}
\usepackage[nameinlink,noabbrev]{cleveref}

\newcommand{\AblationRescued}{13}
\newcommand{\AblationTotal}{98}
\newcommand{\CompressionNinetyNineFifty}{5.25}
\newcommand{\CompressionNinetyNineSeventyFive}{4.78}
\newcommand{\CompressionNinetyNineTwentyFive}{4.14}
\newcommand{\CompressionNinetyNineZero}{3.54}
\newcommand{\CoveragePrunedNinetyNineFifty}{83.6}
\newcommand{\CoveragePrunedNinetyNineSeventyFive}{71.4}
\newcommand{\CoveragePrunedNinetyNineTwentyFive}{85.7}
\newcommand{\CoveragePrunedNinetyNineZero}{85.7}
\newcommand{\CoverageSearchedNinetyNineFifty}{92.9}
\newcommand{\CoverageSearchedNinetyNineSeventyFive}{88.6}
\newcommand{\CoverageSearchedNinetyNineTwentyFive}{94.3}
\newcommand{\CoverageSearchedNinetyNineZero}{95.7}
\newcommand{\CxRatioAdaptiveCINinetyNineFifty}{[0.423, 0.508]}

\newcommand{\CxRatioAdaptiveNNinetyNineFifty}{117}

\newcommand{\CxRatioAdaptiveNinetyNineFifty}{0.464}

\newcommand{\CxRatioCINinetyNineFifty}{[0.946, 0.997]}
\newcommand{\CxRatioCINinetyNineSeventyFive}{[0.984, 1.042]}
\newcommand{\CxRatioCINinetyNineTwentyFive}{[0.977, 1.041]}
\newcommand{\CxRatioCINinetyNineZero}{[0.971, 1.026]}
\newcommand{\CxRatioCommonCINinetyNineFifty}{[0.653, 0.716]}

\newcommand{\CxRatioCommonNNinetyNineFifty}{109}

\newcommand{\CxRatioCommonNinetyNineFifty}{0.684}

\newcommand{\CxRatioExactCINinetyNineFifty}{[0.160, 0.222]}

\newcommand{\CxRatioExactNNinetyNineFifty}{117}

\newcommand{\CxRatioExactNinetyNineFifty}{0.189}

\newcommand{\CxRatioNNinetyNineFifty}{117}
\newcommand{\CxRatioNNinetyNineSeventyFive}{100}
\newcommand{\CxRatioNNinetyNineTwentyFive}{120}
\newcommand{\CxRatioNNinetyNineZero}{120}
\newcommand{\CxRatioNinetyNineFifty}{0.971}
\newcommand{\CxRatioNinetyNineSeventyFive}{1.013}
\newcommand{\CxRatioNinetyNineTwentyFive}{1.009}
\newcommand{\CxRatioNinetyNineZero}{0.998}
\newcommand{\CxRatioVfiveCINinetyNineFifty}{[1.064, 1.140]}

\newcommand{\CxRatioVfiveNNinetyNineFifty}{117}

\newcommand{\CxRatioVfiveNinetyNineFifty}{1.101}
\newcommand{\CxRatioVfiveNinetyNineSeventyFive}{1.146}

\newcommand{\CxRatioVfiveNinetyNineZero}{1.114}
\newcommand{\DataSpecSha}{6bddf4e8d2e3}

\newcommand{\ExpectedPrunedExactNinetyNineFifty}{143.5}
\newcommand{\ExpectedPrunedExactNinetyNineSeventyFive}{204.2}

\newcommand{\ExpectedPrunedExactNinetyNineZero}{121.0}
\newcommand{\ExpectedPrunedSearchedNinetyNineFifty}{113.2}

\newcommand{\ExpectedSearchedExactNinetyNineFifty}{116.0}
\newcommand{\ExpectedSearchedExactNinetyNineSeventyFive}{143.1}

\newcommand{\ExpectedSearchedExactNinetyNineZero}{85.7}
\newcommand{\ExpectedSearchedSearchedNinetyNineFifty}{116.0}

\newcommand{\HifdaVoltageAliasingMax}{97}
\newcommand{\HifdaVoltageAliasingMin}{54}
\newcommand{\HifdaVoltageEnergyAboveOneKilohertz}{97}
\newcommand{\HifdaVoltageFundamentalShare}{1.7}

\newcommand{\MainCases}{140}

\newcommand{\MeanOnlyETest}{0.6480}
\newcommand{\NegativeControlRuns}{28}

\newcommand{\OrderOneETest}{0.6279}
\newcommand{\PairedCompressionNNinetyNineFifty}{117}

\newcommand{\PairedKQnlNinetyNineFifty}{77.2}

\newcommand{\PairedKSearchedNinetyNineFifty}{82.3}

\newcommand{\PairedRQnlNinetyNineFifty}{13.4}

\newcommand{\PruneCxRatioNinetyNineFifty}{0.882}
\newcommand{\PruneCxRatioNinetyNineSeventyFive}{0.884}
\newcommand{\PruneCxRatioNinetyNineTwentyFive}{0.882}
\newcommand{\PruneCxRatioNinetyNineZero}{0.896}
\newcommand{\PruneKRatioNinetyNineFifty}{0.853}

\newcommand{\PrunedCxNinetyNineFifty}{83.5}

\newcommand{\QnlDofSixteenETest}{0.0890}
\newcommand{\QnlDofTwoETest}{0.1657}

\newcommand{\ScalingExponentExact}{1.01}
\newcommand{\ScalingExponentExactCI}{[1.00, 1.01]}
\newcommand{\ScalingExponentPruned}{0.18}
\newcommand{\ScalingExponentPrunedCI}{[0.07, 0.29]}
\newcommand{\ScalingExponentSearched}{0.19}
\newcommand{\ScalingExponentSearchedCI}{[-0.09, 0.47]}
\newcommand{\ScalingRuns}{17}

\newcommand{\ScalingRunsPlanned}{18}
\newcommand{\SearchedCxNinetyNineFifty}{110.3}

\newcommand{\SurrogateAdaptiveDeployed}{28}
\newcommand{\SurrogateAdaptiveTotal}{28}
\newcommand{\SurrogatePrunedDeployed}{0}
\newcommand{\SurrogatePrunedTotal}{28}

\newcommand{\UnitStressCoveragePruned}{84.6}
\newcommand{\UnitStressRecordingCoveragePruned}{82.9}
\newcommand{\UnitStressRecordingRuns}{35}
\newcommand{\UnitStressRuns}{39}

\newcommand{\WalshStaticDofSixteenETest}{0.3077}
\newcommand{\WalshStaticDofTwoETest}{0.5764}

\newcommand{\BitIdenticalCells}{35364}
\newcommand{\ComparedCells}{35364}
\newcommand{\DeployedPrunedModels}{457}

\newcommand{\PartialQnlCasesMeetingNinetyNineFiftyObsThirty}{69}

\newcommand{\PartialQnlNNinetyNineFiftyObsHundred}{116}

\newcommand{\PartialQnlNNinetyNineFiftyObsThirty}{116}

\newcommand{\PartialQnlNinetyNineFiftyObsHundred}{0.0622}

\newcommand{\PartialQnlNinetyNineFiftyObsTwenty}{0.0734}

\newcommand{\PartialQnlUnknownsNinetyNineFiftyObsHundred}{13.3}

\newcommand{\PartialQnlVfiveNinetyNineFiftyObsHundred}{0.0563}

\newcommand{\PartialQnlVfiveNinetyNineFiftyObsTwenty}{0.0661}

\newcommand{\PartialWalshMatchedKNinetyNineFiftyObsHundred}{0.3261}

\newcommand{\PartialWalshMatchedKUnknownsNinetyNineFiftyObsHundred}{13.3}

\newcommand{\PartialWalshStaticUnknownsNinetyNineFiftyObsHundred}{82.6}

\newcommand{\RankCheckedModels}{453}
\newcommand{\RankDeficientModels}{4}
\newcommand{\ReproducedPrunedModels}{453}

\newcommand{\BandHigh}{1.05}
\newcommand{\BandLow}{0.95}
\newcommand{\BasisGates}{\texttt{rz}, \texttt{sx}, \texttt{x} and \texttt{cx}}
\newcommand{\BlockTolerance}{$10^{-6}$}
\newcommand{\CentralFidelity}{99.50\%}
\newcommand{\ContractConfidence}{95\%}
\newcommand{\DecoderRidge}{$10^{-8}$}
\newcommand{\DecoderRounds}{3}
\newcommand{\FinalProtocolSha}{447826a5e1d9}

\newcommand{\GateErrors}{$10^{-4}$, $5\times10^{-4}$, $10^{-3}$ and $5\times10^{-3}$}
\newcommand{\MainDimensions}{128, 256, 512 and 1024}
\newcommand{\MaxSupportProposals}{64}
\newcommand{\MinReplicates}{3}
\newcommand{\NativeLength}{4096}
\newcommand{\NativeLengthPtbxl}{1024}
\newcommand{\NegativeControlDimension}{256}
\newcommand{\NegativeControlSeed}{0}
\newcommand{\ObservedFractions}{1.00, 0.50, 0.30, 0.20, 0.10 and 0.05}
\newcommand{\OptimizationLevel}{3}
\newcommand{\PartialProtocolSha}{df0916641834}

\newcommand{\PartialRcond}{$10^{-10}$}

\newcommand{\PhaseMargin}{0.95}
\newcommand{\PhaseQuantile}{0.99}
\newcommand{\PruneMinFraction}{0.5}
\newcommand{\PruneShortlist}{8}
\newcommand{\PruneStep}{8}
\newcommand{\RGrid}{2, 4, 8 and 16}
\newcommand{\RGridAblation}{32 and 64}
\newcommand{\ScalingDimension}{2048}
\newcommand{\ScalingSeeds}{0, 1 and 2}
\newcommand{\SeedCount}{5}
\newcommand{\SplitShares}{70/20/10}
\newcommand{\TargetFidelities}{99.00\%, 99.25\%, 99.50\% and 99.75\%}
\newcommand{\TranspileSeed}{7}
\newcommand{\UnitFoldSeed}{0}
\newcommand{\UnitFoldsHifda}{6}
\newcommand{\UnitFoldsPaderborn}{8}
\newcommand{\UnitSplitMinUnits}{50}
\newcommand{\UnitStressDimension}{512}

\newcommand{\ViewDimensions}{128, 256, 512, 1024 and 2048}

\newcommand{\ObsVthreeAbsSlopePruned}{0.49}
\newcommand{\ObsVthreeAbsSlopePrunedCI}{[0.39, 0.58]}

\newcommand{\ObsVthreeAxisFirst}{16}
\newcommand{\ObsVthreeAxisLast}{2048}
\newcommand{\ObsVthreeAxisPoints}{15}

\newcommand{\ObsVthreeContiguousSlopePruned}{0.92}

\newcommand{\ObsVthreeContiguousSlopeSearched}{1.00}
\newcommand{\ObsVthreeDelta}{10}

\newcommand{\ObsVthreeDeltaFiveSlopePruned}{0.22}
\newcommand{\ObsVthreeDeltaFiveSlopePrunedCI}{[0.18, 0.25]}

\newcommand{\ObsVthreeDimensionFirst}{128}
\newcommand{\ObsVthreeDimensionLast}{2048}
\newcommand{\ObsVthreeEquivalenceMargin}{0.15}

\newcommand{\ObsVthreeMeanLatent}{13.5}
\newcommand{\ObsVthreeMinDimensionsPerUnit}{3}
\newcommand{\ObsVthreeOriginSeedLast}{14}

\newcommand{\ObsVthreePerCoordinateMax}{6.4}
\newcommand{\ObsVthreePerCoordinateMin}{4.9}

\newcommand{\ObsVthreeRelGeoPrunedDOneTwoEight}{61}

\newcommand{\ObsVthreeRelGeoPrunedDTwoZeroFourEight}{76}

\newcommand{\ObsVthreeRelGeoSearchedDOneTwoEight}{116}

\newcommand{\ObsVthreeRelGeoSearchedDTwoZeroFourEight}{464}

\newcommand{\ObsVthreeRelSharePrunedDOneTwoEight}{47.3}

\newcommand{\ObsVthreeRelSharePrunedDTwoZeroFourEight}{3.7}

\newcommand{\ObsVthreeSeedFirst}{15}
\newcommand{\ObsVthreeSeedLast}{19}
\newcommand{\ObsVthreeSlopeContrast}{0.43}
\newcommand{\ObsVthreeSlopeContrastCI}{[0.35, 0.50]}
\newcommand{\ObsVthreeSlopePruned}{0.08}
\newcommand{\ObsVthreeSlopePrunedCI}{[0.05, 0.12]}
\newcommand{\ObsVthreeSlopeSearched}{0.51}
\newcommand{\ObsVthreeSlopeSearchedCI}{[0.44, 0.58]}
\newcommand{\ObsVthreeTestRatioPrunedMax}{1.09}
\newcommand{\ObsVthreeTestRatioPrunedMin}{1.08}

\newcommand{\ObsVthreeVfiveSlope}{0.08}

\newcommand{\RepositoryURL}{\url{https://github.com/phg-tec/quantum-native-loader-public}}
\newcommand{\ArchiveDOI}{\url{https://doi.org/10.5281/zenodo.23104574}}

\title{Quantum data loading from the learned shared structure of real signals}
\author[1,*]{Pablo Herrero G\'omez}
\author[1]{Antonio Jimeno Morenilla}
\author[1]{David Mu\~noz-Hern\'andez}
\author[1]{Higinio Mora Mora}

\affil[1]{Department of Computer Science Technology and Computation,
Universidad de Alicante, San Vicente del Raspeig, 03690 Alicante, Spain}

\affil[*]{Corresponding author: \texttt{pablo.herrero@ua.es}}
\date{\today}

\begin{document}

\maketitle
\begin{abstract}
Preparing quantum states from classical data can cost more than the computation they serve; most
loaders tailor a circuit to each input. Here we show that the signals of a real dataset share
structure that can be learned once and reused. Our quantum-native loader learns a low-dimensional
description of a dataset and prepares every signal with one fixed circuit set by a few numbers.
Across seven views of five public datasets it meets the targets of the strongest structured loader
at equal gate cost with several times fewer numbers per signal. These numbers can be inferred from
a random subset: in a preregistered blind replication the subset needed to come within ten per cent
of full-signal accuracy stayed constant within a prespecified margin as signals grew sixteenfold,
whereas the structured loader needed ever more. It declines what it cannot represent, covering
fewer cases than that baseline and no electrocardiogram.
\end{abstract}

\section{Introduction}
\label{sec:introduction}

Quantum algorithms that act on classical data must first represent that data in a quantum
register. This input step is easy to hide in an oracle assumption, but its cost can remove a
downstream speed-up when the state must be prepared explicitly
\cite{aaronson2015fineprint,biamonte2017qml}. Quantum random-access memory provides one access
model\cite{giovannetti2008qram}, but is not available as a general primitive on present devices.
In gate-based implementations, state preparation therefore contributes directly to circuit depth,
two-qubit gate count and accumulated error\cite{preskill2018nisq}. That bound is a property of
the encoding, not of the hardware\cite{faryad2026inputproblem}, so surviving advantage turns on
data access\cite{zhao2026massive}.

The worst-case cost is unavoidable. Preparing an arbitrary state of $n$ qubits requires a number
of two-qubit gates that grows linearly with the number of amplitudes and hence exponentially with
$n$\cite{grover2002superpositions,mottonen2005uniformly,shende2006synthesis,plesch2011universal}.
Ancillas can trade gate count for depth but do not remove the dependence on an arbitrary target
description\cite{zhang2022optimaldepth,sun2023asymptotically,araujo2021divide}. The same issue
occurs for the phase encoding studied here: a window of $D=2^n$ real values specifies a diagonal
unitary, and an arbitrary exact diagonal requires resources that grow with
$D$\cite{bullock2004diagonal,welch2014diagonal}.

Such a state is what a diagonal potential, phase oracle or signal-processing transform
consumes\cite{low2017qsp,gilyen2019qsvt,motlagh2024gqsp,sarmina2026qpipe}.

Practical data are not arbitrary, and structured state preparation exploits this fact. Truncated
Walsh and Fourier representations retain a small number of
coefficients\cite{zylberman2024walsh,moosa2024fsl}; smooth, polynomial and multivariate functions
admit specialized constructions
\cite{holmes2020smooth,gonzalezconde2024polynomial,marinsanchez2023approximate,rosenkranz2025multivariate};
quantum signal-processing constructions prepare classes of continuous or piecewise-defined
functions\cite{mcardle2022arithmetic,obrien2025piecewise,rattew2022continuous}; and tensor-network
factorizations expose low-entanglement structure in images, financial data and physical
fields\cite{ran2020mps,dilip2022compression,jobst2024mps,jumade2023shortdepth}.
Approximation-aware compilation and entanglement-reduction frameworks automate the choice of a
resource-efficient construction for a supplied
target\cite{alonsolinaje2025compilation,zhang2026aqer}. In these target-specific approaches, some
combination of retained support, factorization, approximation or circuit structure is determined
from the complete input instance.

Shared structure across targets has also begun to be used explicitly. Daimon and Matsushita train a
transformer decoder to generate gate sequences for previously unseen amplitude
vectors\cite{daimon2024transformer}. Aoun et al. instead map each complete input image to the
continuous parameters of a fixed circuit topology\cite{aoun2026neural}. The adaptive interpolating
quantum transform learns a shared, circuit-realizable basis across a dataset, followed at
deployment by a full classical transform and target-specific top-$k$ coefficient
selection\cite{budiutama2026aiqt}, and dataset distillation amortizes whole datasets into shallow
circuits\cite{lin2026distilling}. These works establish that circuit
generation, parameter prediction or basis design can be amortized across inputs. Offline learning
and reuse are therefore not by themselves the distinguishing question.

Even when circuit design is amortized, a separate input question remains: how much information
from each new instance must be acquired and processed to instantiate its quantum state? A model
that consumes the complete vector can avoid per-input circuit optimization while retaining a
classical interface of dimension $D$. Conversely, a lower-dimensional interface could reduce the
number of per-instance unknowns and might allow those unknowns to be inferred when some positions
are unobserved. Such a claim must be distributional: it can hold for new members of a learned
family, not for an arbitrary vector, and it requires complete examples during learning.

We investigate the hypothesis that a source of real signals occupies a low-dimensional family that
is simultaneously statistical and physical. In QNL, the physical coordinates are coefficients on
a shared Walsh support $S$. Offline, on complete FIT windows, the support is searched and pruned with
measured transpiled CX cost and an affine decoder $c(z)=b+Az$ is fitted; VALIDATION selects models
under a statistical deployment contract, and the selected model and circuit topology are frozen. The
nonlinear families used to initialize FIT coordinates and propose supports are not the deployed
loader and never see TEST windows.

Online, a new window is not assigned a newly optimized circuit: its coordinates are obtained by
linear least-squares projection onto the frozen family, and $c=b+A\hat z$ supplies the coefficients
of the fixed Walsh-gadget circuit. If only some positions are available, the same projection uses
the corresponding rows, and no complete classical reconstruction is needed to obtain the circuit
parameters. We test this on seven views of five public datasets, keeping circuit CX, per-instance
interface dimension, observed positions, offline learning and online inference as distinct
resources. The central result concerns how many positions must be observed. In a preregistered blind
replication with random observations, the number of positions QNL needs to approach its full-window
accuracy (an error within \ObsVthreeDelta\% of that with the whole window) was constant within a
prespecified margin from 128 to 2048 samples, with a slope on logarithmic axes of
\ObsVthreeSlopePruned{} (90\% interval \ObsVthreeSlopePrunedCI{}), whereas the strongest shared Walsh
baseline's grew with slope \ObsVthreeSlopeSearched{}; it is not the number needed to meet the
fidelity budget, and with a contiguous gap there is no advantage. Each window supplies
\CompressionNinetyNineFifty{} times fewer numbers to QNL than to that baseline, although both
circuits receive one coefficient per support term, and the two circuits reach paired CX parity. QNL
abstains when its validation contract is not met, and its lower coverage, particularly on PTB-XL,
delineates the empirical scope of the affine family.

\section{Results}
\label{sec:results}

\begin{figure}[tbp]
  \centering
  \includegraphics[width=\textwidth]{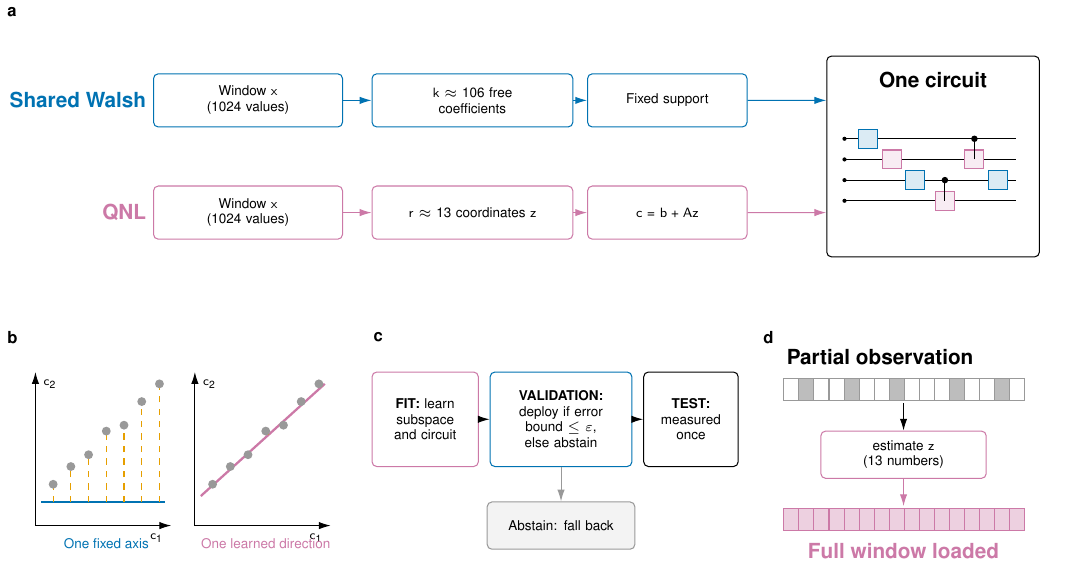}
  \caption{\textbf{A learned family turns a window into a few coordinates of one fixed circuit.}
  \textbf{a}, Loading pipeline. A shared Walsh support (top) takes $k$ free coefficients from the
  complete window; QNL (bottom) takes $r$ latent coordinates $z$ and evaluates the coefficients
  through the affine decoder $c=b+Az$. Both paths end in one circuit whose topology is fixed for
  the whole dataset and whose phase gadgets differ between windows only in their angles, so both
  circuits receive one coefficient per support term: what differs is how many numbers each window
  has to supply. The counts shown are means over the deployed models at the central target
  fidelity (99.50\%).
  \textbf{b}, Why that interface is narrower. The same seven windows, in the plane of two Walsh
  coefficients. One fixed axis leaves a large residual on every window (left, dashed); one
  direction fitted to the family leaves a small one (right), so fewer coordinates describe the
  same windows. The fitted directions are constrained to lie in a physically realizable Walsh
  support, which is what distinguishes them from an unconstrained principal
  component\cite{jolliffe2002pca}.
  \textbf{c}, Protocol. The subspace, the support and the circuit are learned on FIT;
  VALIDATION applies the deployment contract and the method abstains, and is priced at a declared
  fallback, when that contract is not met; TEST is materialised once, after every selection has
  been frozen.
  \textbf{d}, Partial observation. From the observed positions alone the least-squares projection
  returns the $r$ coordinates, and these already determine the circuit angles through
  $c=b+A\hat z$; the complete window is reconstructed only to score the error against the held-out
  window, never as a deployment object.}
  \label{fig:method}
\end{figure}

All results come from one frozen evaluation protocol (Supplementary Note~1; Methods), summarised
with the loader in Fig.~\ref{fig:method}. A window of
a real signal was loaded as the phase of a quantum state and scored by
$E_{\mathrm{state}}=\sqrt{1-F}$ on held-out TEST windows. A model was deployed at a target fidelity
$F$ only if the one-sided 95\% upper confidence limit of $E_{\mathrm{state}}$ on VALIDATION,
clustered by recording, was at most $\epsilon=\sqrt{1-F}$ and its circuit used fewer two-qubit gates
than the exact preparation; otherwise the method abstained, and the abstention counts against it.
Four target fidelities were evaluated, 99.00\%, 99.25\%, 99.50\% and 99.75\%, never pooled; where
one is needed it is the central fidelity, 99.50\%. The main campaign comprised \MainCases{} cases:
seven views of five public
datasets\cite{medico2020plaid,kahl2016whited,lessmeier2016paderborn,navarro2025hifda,wagner2020ptbxl},
four window lengths $D$ from 128 to 1024, and five seeds (Supplementary Note~2).

Three quantities are kept apart: $r$, the number of sample-specific numbers the loader needs per
window; $K$, the number of Walsh terms in the physical support, and therefore the number of
coefficients the circuit receives per window; and CX, the two-qubit gates after transpilation.
QNL-pruned learns on FIT a shared affine decoder $c(z)=b+Az$ from the $r$ coordinates of a window to
the coefficients of a shared Walsh support, prunes that support with a refit after each removal,
and deploys the candidate with the fewest two-qubit gates that meets the contract on VALIDATION;
QNL-v5, the same model before pruning, is an ablation. Once selected, the model and its circuit are
frozen, and a new window needs only a least-squares projection onto the fixed subspace followed by
$c=b+A\hat z$ (Methods).

\subsection{A learned family needs fewer numbers per window}
\label{sec:representation}

We first compared methods at equal cost in sample-specific numbers, without an error budget.
QNL was given $r=2,4,8,16$ coordinates per window and each shared Walsh baseline the same number
of free coefficients, $k=r$; QNL was not pruned in this comparison, because pruning trades
accuracy for gates and belongs to the physical comparison that follows.

At every matched count, QNL reaches a lower TEST error than any Walsh baseline given the same
number of free coefficients per window (Fig.~\ref{fig:representation}a; per dataset in Supplementary Figure~1). With two numbers per
window its error is \QnlDofTwoETest{}, and with sixteen it is \QnlDofSixteenETest{}; the shared
static Walsh support at the same two counts reaches \WalshStaticDofTwoETest{} and
\WalshStaticDofSixteenETest{}. The two trivial references bound the axis from below:
the FIT template alone, with no number per window, reaches \MeanOnlyETest{}, and the template
with the $\log_2 D$ single-qubit Walsh masks reaches \OrderOneETest{}. The exact diagonal
preparation\cite{bullock2004diagonal} has no representation error but needs $D-1$ numbers per
window (Supplementary Table~1).

The same saving appears when every method is instead held to a target fidelity and left free to
choose its own size. For each case where both QNL-pruned and the searched shared Walsh support
deployed, we took the ratio of the baseline's free coefficients to QNL's $r$
(Fig.~\ref{fig:representation}b). At the central fidelity its geometric mean is
\CompressionNinetyNineFifty{}; at 99.00\%, 99.25\% and 99.75\% it is \CompressionNinetyNineZero{},
\CompressionNinetyNineTwentyFive{} and \CompressionNinetyNineSeventyFive{} (Supplementary
Table~2). The ratio varies markedly between datasets.

This comparison isolates the dimension of the per-instance interface. A subspace oriented to the
observed data can describe the same family with fewer coordinates than a fixed set of axes, as in
principal component analysis\cite{jolliffe2002pca}; that geometric fact is not itself new. Here the
learned directions are constrained to a physically realizable Walsh support. Offline learning fixes $S$, $b$ and $A$; online inference estimates $r$ coordinates and evaluates $K$ physical coefficients as $c=b+A\hat z$. The factor \CompressionNinetyNineFifty{} therefore measures how many numbers each window has to supply, not how many the circuit receives: on the same \PairedCompressionNNinetyNineFifty{} pairs QNL-pruned estimates \PairedRQnlNinetyNineFifty{} coordinates per window on average, but its circuit takes \PairedKQnlNinetyNineFifty{} coefficients, against \PairedKSearchedNinetyNineFifty{} for the baseline's (Supplementary Note~3). It is not a reduction in the data transferred to the hardware, and no runtime was measured. The smaller number of online unknowns is also what
makes estimation from incomplete observations possible.

\begin{figure}[tbp]
  \centering
  \includegraphics[width=\textwidth]{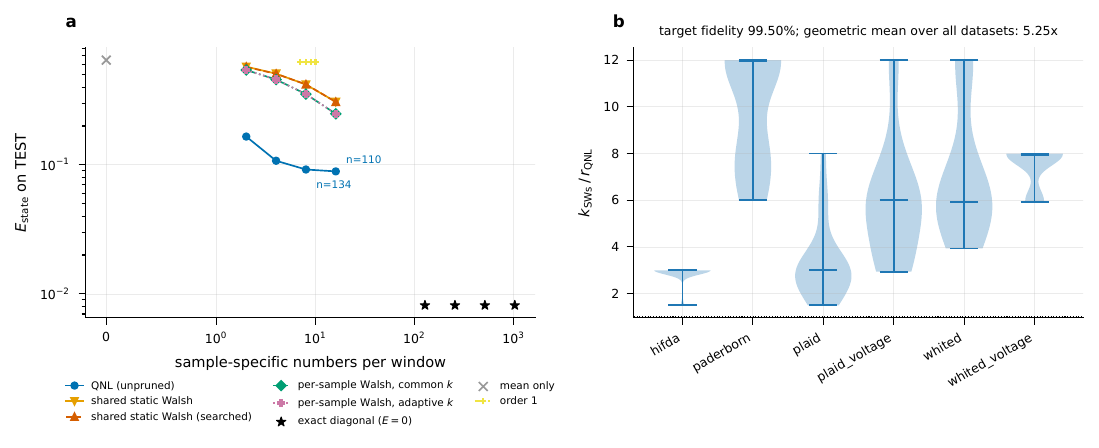}
  \caption{\textbf{One shared decoder describes a window with a fraction of the numbers a Walsh
  support needs.}
  \textbf{a}, State-preparation error $E_{\mathrm{state}}=\sqrt{1-F}$ on TEST against the number
  of sample-specific degrees of freedom exposed by each method, with no target fidelity imposed.
  Points are means over the cases of the main campaign that have a model at that count; no error
  bars are drawn. Where QNL has fewer cases than the baselines its $n$ is printed beside the
  point, because a case whose selected latent is smaller cannot be evaluated at a larger one. The
  exact diagonal has zero error and is drawn on the lower edge at $D-1$ numbers.
  \textbf{b}, Ratio $k_{\mathrm{SWs}}/r_{\mathrm{QNL}}$ of the free coefficients the searched shared
  Walsh support needs to QNL-pruned's $r$, per dataset, at the central target fidelity (99.50\%),
  over the cases where both deploy; violins show the distribution over cases and seeds, bars the
  per-dataset median, and the title the geometric mean over all datasets
  (\CompressionNinetyNineFifty{}). PTB-XL is absent because QNL-pruned deploys on none of its
  cases (Fig.~\ref{fig:limits}a).}
  \label{fig:representation}
\end{figure}

\subsection{Gate cost ties with the strongest Walsh baseline}
\label{sec:cost}

\begin{figure}[tbp]
  \centering
  \includegraphics[width=\textwidth]{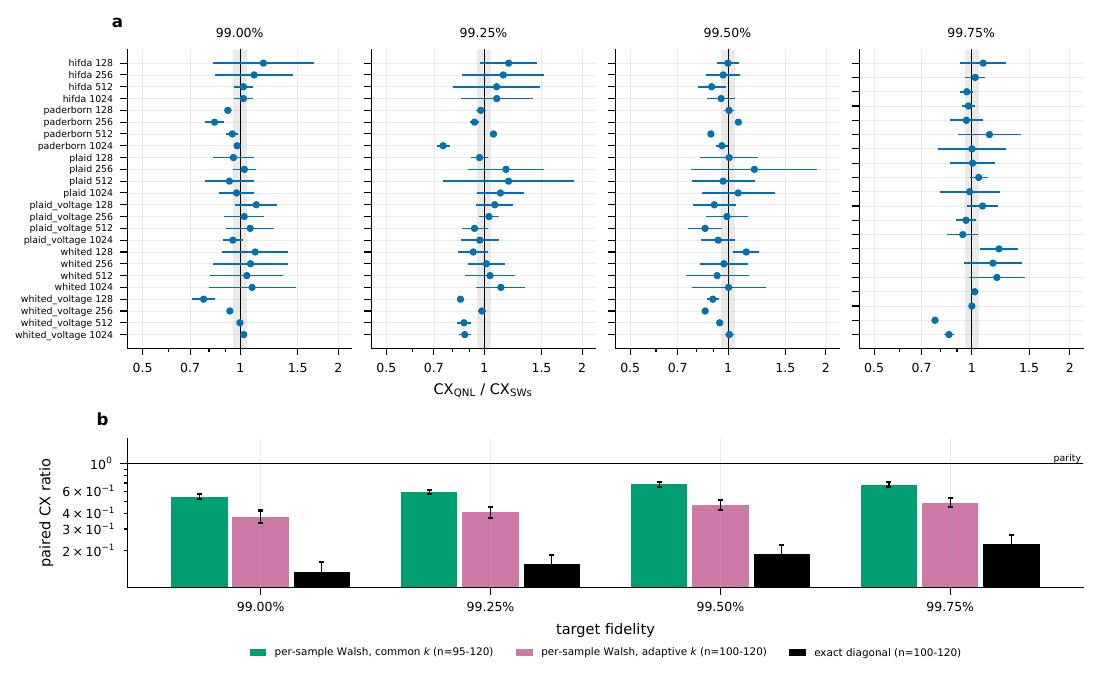}
  \caption{\textbf{The circuit is no more expensive than the strongest Walsh baseline and far
  cheaper than per-sample and exact loaders.}
  \textbf{a}, Paired ratio of two-qubit gate count, QNL-pruned over the searched shared Walsh
  support, one row per dataset and window length $D$, one panel per target fidelity. Points are
  geometric means over the seeds in which both methods deploy; bars are 95\% Student intervals
  taken on the log scale. Cells with fewer than three paired cases are drawn as open points
  without an interval. The shaded band is the declared equivalence band $[0.95, 1.05]$ and the
  line is parity. PTB-XL has no row because QNL-pruned deploys on none of its cases.
  \textbf{b}, Paired CX ratio of QNL-pruned to the per-sample Walsh support with a common $k$,
  with an adaptive $k$, and to the exact diagonal, per target fidelity; bars are geometric means
  over paired cases and whiskers 95\% intervals on the log scale. The range of $n$ over the four
  fidelities is given in the legend. Aggregates are never pooled across fidelities.}
  \label{fig:cost}
\end{figure}

Fewer numbers per window would be of little use if the circuit that consumes them were more
expensive. We therefore compared the two-qubit gate count each method needs to meet a target
fidelity, pairing methods within a case and a fidelity so that each ratio compares the same
window population (Fig.~\ref{fig:cost}, Table~\ref{tab:resources}). The strongest physical
baseline is the shared static Walsh support with a searched mask
set\cite{welch2014diagonal,zylberman2024walsh}.

Against it, QNL-pruned is at parity. At the central fidelity the paired CX ratio is
\CxRatioNinetyNineFifty{} (95\% interval \CxRatioCINinetyNineFifty{}, $n=\CxRatioNNinetyNineFifty{}$
paired cases), and at 99.00\%, 99.25\% and 99.75\% it is \CxRatioNinetyNineZero{}
(\CxRatioCINinetyNineZero{}, $n=\CxRatioNNinetyNineZero{}$), \CxRatioNinetyNineTwentyFive{}
(\CxRatioCINinetyNineTwentyFive{}, $n=\CxRatioNNinetyNineTwentyFive{}$) and
\CxRatioNinetyNineSeventyFive{} (\CxRatioCINinetyNineSeventyFive{},
$n=\CxRatioNNinetyNineSeventyFive{}$). All four geometric means lie inside the declared
equivalence band $[0.95, 1.05]$, which the protocol reads as a tie: on the interval, equivalence
at three targets and non-inferiority at the central one, below the band on the cheaper side.
Across datasets and window lengths the per-cell
ratios scatter on both sides of one (Fig.~\ref{fig:cost}a, Supplementary Table~3; against every baseline in Supplementary Figure~2). The unpaired
means in Table~\ref{tab:resources} (\PrunedCxNinetyNineFifty{} against \SearchedCxNinetyNineFifty{}
at the central fidelity) are not this comparison: they are taken over different sets of cases,
because the baseline also deploys on cases where QNL-pruned abstains, and those are among the
most expensive.

Against the loaders that adapt to each window, the saving is large at every fidelity. The paired
CX ratio of QNL-pruned to the per-sample Walsh support with a common $k$, to the per-sample
support with an adaptive $k$ and to the exact diagonal is below one at all four targets
(Fig.~\ref{fig:cost}b). At the central fidelity it is \CxRatioCommonNinetyNineFifty{}
(\CxRatioCommonCINinetyNineFifty{}, $n=\CxRatioCommonNNinetyNineFifty{}$) against the common-$k$
support, \CxRatioAdaptiveNinetyNineFifty{} (\CxRatioAdaptiveCINinetyNineFifty{},
$n=\CxRatioAdaptiveNNinetyNineFifty{}$) against the adaptive-$k$ support and
\CxRatioExactNinetyNineFifty{} (\CxRatioExactCINinetyNineFifty{},
$n=\CxRatioExactNNinetyNineFifty{}$) against the exact diagonal (Supplementary Table~3). These baselines spend their gates on a support
chosen per window, and the exact diagonal on all $D-1$ terms.

Pruning is what brings QNL to parity. Starting from the support QNL-v5 selects, removing masks
one at a time with a refit after each removal reduced $K$ by a paired factor of
\PruneKRatioNinetyNineFifty{} and CX by \PruneCxRatioNinetyNineFifty{} at the central fidelity; at
99.00\%, 99.25\% and 99.75\% the CX factor is \PruneCxRatioNinetyNineZero{},
\PruneCxRatioNinetyNineTwentyFive{} and \PruneCxRatioNinetyNineSeventyFive{}. The gap between TEST
and VALIDATION error did not widen: it is of the same size for QNL-pruned and QNL-v5 at every
fidelity (Supplementary Table~4). The pruned support never admits a mask outside the one QNL-v5
selected, so this is a saving in gates, not a different model class. Before pruning, QNL-v5 is
outside the equivalence band against the searched baseline at every fidelity, with a paired CX
ratio of \CxRatioVfiveNinetyNineFifty{} (\CxRatioVfiveCINinetyNineFifty{},
$n=\CxRatioVfiveNNinetyNineFifty{}$) at the central fidelity and between
\CxRatioVfiveNinetyNineZero{} and \CxRatioVfiveNinetyNineSeventyFive{} at the other three.

\begin{table}[tbp]
  \centering
  \caption{\textbf{Resources per target fidelity.} For each method, the number of the
  \MainCases{} cases where it deployed, and the mean two-qubit gate count (CX), circuit depth,
  number of physical Walsh terms ($K$), sample-specific numbers per window and TEST fidelity
  over the cases where it deployed. Means are taken over each method's own deployed cases and
  are therefore not paired comparisons; coverage is the first column. \textit{mean only} and
  \textit{order 1} deploy on no case.}
  \label{tab:resources}
  \footnotesize
  \setlength{\tabcolsep}{2.2pt}
  \begin{tabular}{llllllll}
\toprule
target fidelity & method & deployed & CX & depth & $K$ & numbers per window & TEST fidelity \\
\midrule
99.00\% & QNL-pruned & 120/140 & 61.4 & 109.8 & 57.6 & 13.8 & 0.98675 \\
99.00\% & QNL-v5 (ablation) & 120/140 & 69.4 & 123.2 & 66.0 & 13.8 & 0.98810 \\
99.00\% & mean only & 0/140 & -- & -- & -- & -- & -- \\
99.00\% & order 1 & 0/140 & -- & -- & -- & -- & -- \\
99.00\% & shared static Walsh & 134/140 & 85.4 & 155.8 & 79.6 & 79.6 & 0.98988 \\
99.00\% & shared static Walsh (searched) & 134/140 & 83.0 & 152.3 & 79.2 & 79.2 & 0.98901 \\
99.00\% & per-sample Walsh, common $k$ & 135/140 & 128.2 & 196.6 & 77.7 & 77.7 & 0.99309 \\
99.00\% & per-sample Walsh, adaptive $k$ & 140/140 & 180.5 & 307.1 & 60.2 & 60.2 & 0.99476 \\
99.00\% & exact diagonal & 140/140 & 478.0 & 943.0 & 479.0 & 479.0 & 1.00000 \\
99.25\% & QNL-pruned & 120/140 & 69.0 & 124.1 & 64.3 & 13.3 & 0.98999 \\
99.25\% & QNL-v5 (ablation) & 120/140 & 81.2 & 145.0 & 77.3 & 13.3 & 0.99139 \\
99.25\% & mean only & 0/140 & -- & -- & -- & -- & -- \\
99.25\% & order 1 & 0/140 & -- & -- & -- & -- & -- \\
99.25\% & shared static Walsh & 132/140 & 93.6 & 170.8 & 87.4 & 87.4 & 0.99196 \\
99.25\% & shared static Walsh (searched) & 132/140 & 91.4 & 167.0 & 87.1 & 87.1 & 0.99154 \\
99.25\% & per-sample Walsh, common $k$ & 126/140 & 132.6 & 203.9 & 80.6 & 80.6 & 0.99422 \\
99.25\% & per-sample Walsh, adaptive $k$ & 140/140 & 190.8 & 326.0 & 68.0 & 68.0 & 0.99563 \\
99.25\% & exact diagonal & 140/140 & 478.0 & 943.0 & 479.0 & 479.0 & 1.00000 \\
99.50\% & QNL-pruned & 117/140 & 83.5 & 150.0 & 77.2 & 13.4 & 0.99293 \\
99.50\% & QNL-v5 (ablation) & 117/140 & 95.4 & 170.8 & 91.0 & 13.4 & 0.99394 \\
99.50\% & mean only & 0/140 & -- & -- & -- & -- & -- \\
99.50\% & order 1 & 0/140 & -- & -- & -- & -- & -- \\
99.50\% & shared static Walsh & 130/140 & 112.5 & 205.9 & 105.8 & 105.8 & 0.99522 \\
99.50\% & shared static Walsh (searched) & 130/140 & 110.3 & 202.4 & 105.4 & 105.4 & 0.99509 \\
99.50\% & per-sample Walsh, common $k$ & 120/140 & 135.1 & 207.2 & 81.6 & 81.6 & 0.99554 \\
99.50\% & per-sample Walsh, adaptive $k$ & 140/140 & 203.3 & 344.5 & 69.5 & 69.5 & 0.99645 \\
99.50\% & exact diagonal & 140/140 & 478.0 & 943.0 & 479.0 & 479.0 & 1.00000 \\
99.75\% & QNL-pruned & 100/140 & 94.6 & 170.2 & 88.4 & 16.0 & 0.99584 \\
99.75\% & QNL-v5 (ablation) & 100/140 & 109.5 & 195.2 & 104.7 & 16.0 & 0.99626 \\
99.75\% & mean only & 0/140 & -- & -- & -- & -- & -- \\
99.75\% & order 1 & 0/140 & -- & -- & -- & -- & -- \\
99.75\% & shared static Walsh & 124/140 & 140.0 & 258.2 & 131.8 & 131.8 & 0.99734 \\
99.75\% & shared static Walsh (searched) & 124/140 & 137.1 & 253.5 & 131.4 & 131.4 & 0.99722 \\
99.75\% & per-sample Walsh, common $k$ & 119/140 & 185.6 & 294.5 & 120.5 & 120.5 & 0.99820 \\
99.75\% & per-sample Walsh, adaptive $k$ & 140/140 & 242.1 & 418.8 & 92.3 & 92.3 & 0.99732 \\
99.75\% & exact diagonal & 140/140 & 478.0 & 943.0 & 479.0 & 479.0 & 1.00000 \\
\bottomrule
\end{tabular}

\end{table}

The tie extends to resolution and to noise. To see how the cost grows with resolution, we added $D=2048$ at the central fidelity for every
view whose native window allows it, and fitted $\mathrm{CX}\propto D^{\alpha}$ to the geometric
mean gate count over deployed cases from $D=128$ to 2048 (Fig.~\ref{fig:scaling}a; per dataset in Supplementary Figure~3). The exact
diagonal grows linearly, with $\alpha=\ScalingExponentExact{}$ (95\% interval
\ScalingExponentExactCI{}), as it must. QNL-pruned grows with $\alpha=\ScalingExponentPruned{}$
(\ScalingExponentPrunedCI{}) and the searched Walsh support with
$\alpha=\ScalingExponentSearched{}$ (\ScalingExponentSearchedCI{}): sublinear growth is a property
of loading through a shared support, which both methods share, rather than an advantage of QNL
over the baseline. Both intervals are wide, because each fit rests on five window lengths. The $D=2048$ point rests on
\ScalingRuns{} of \ScalingRunsPlanned{} planned runs: in the remaining one the frozen source model
could not be reproduced and the cell is reported as lost, not dropped.

The value of a lower gate count on hardware depends on how errors accumulate. We estimated the
total fidelity at the central target as $F_{\mathrm{total}}(p)=(1-E_{\mathrm{test}}^{2})(1-p)^{\mathrm{CX}}$
for two-qubit gate error $p$ (Fig.~\ref{fig:scaling}b). This is an analytic estimate under
independent depolarising error per two-qubit gate, not a simulation: it ignores idling,
crosstalk and readout, and ranks methods by gate count at fixed representation error. The three
methods are compared over the same cases, those on which all of them deploy. On those cases
QNL-pruned and the searched Walsh support are indistinguishable under this estimate, as their
paired gate counts are (Fig.~\ref{fig:cost}a), and both retain most of their fidelity at gate
errors where the exact diagonal has lost most of its own. Under noise, as in gate count, the two
shared-support loaders are tied and both are far ahead of exact loading.

\begin{figure}[tbp]
  \centering
  \includegraphics[width=\textwidth]{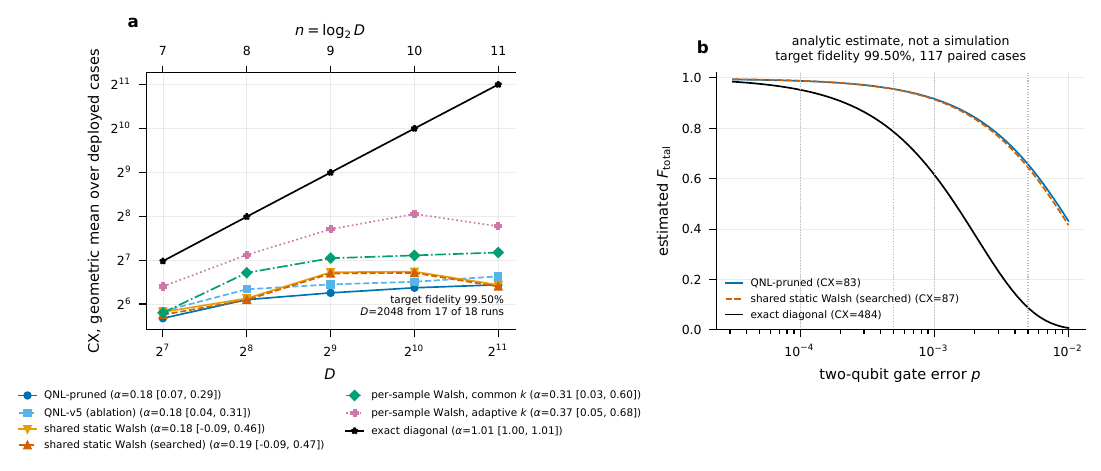}
  \caption{\textbf{Shared supports scale sublinearly with window length, and their saving over
  exact loading survives gate noise.}
  \textbf{a}, Two-qubit gate count against window length $D$, both axes logarithmic, at the
  central target fidelity (99.50\%); points are geometric means over each method's deployed cases
  and the upper axis gives the qubit count $n=\log_2 D$. The fitted exponent $\alpha$ of
  $\mathrm{CX}\propto D^{\alpha}$ and its 95\% interval are given in the legend for each method.
  The $D=2048$ point uses \ScalingRuns{} of \ScalingRunsPlanned{} runs; PTB-XL has no $D=2048$ view.
  \textbf{b}, Estimated total fidelity against two-qubit gate error $p$ at the central target,
  for QNL-pruned, the searched shared Walsh support and the exact diagonal, using each method's
  mean gate count and mean TEST error over the same cases, those where all three deploy; their
  number is given in the panel title and each method's mean CX in the legend.
  \emph{This panel is an analytic estimate, not a simulation.} Dotted lines mark the gate errors
  declared by the protocol.}
  \label{fig:scaling}
\end{figure}

\subsection{Positions needed are constant within a prespecified margin}
\label{sec:partial}

With only the positions $\Omega$ of a window observed, QNL estimates its coordinates from the
corresponding rows, $\hat z=\arg\min_z\lVert x_\Omega-m_\Omega-B_\Omega z\rVert_2$, and
$c=b+A\hat z$ gives the coefficients of the same fixed circuit; nothing is refitted, and the
complete window is used only to score the error. A shared Walsh support estimates its $k$
coefficients in the same way (Methods, Supplementary Note~4).

\begin{figure}[tbp]
  \centering
  \includegraphics[width=\textwidth]{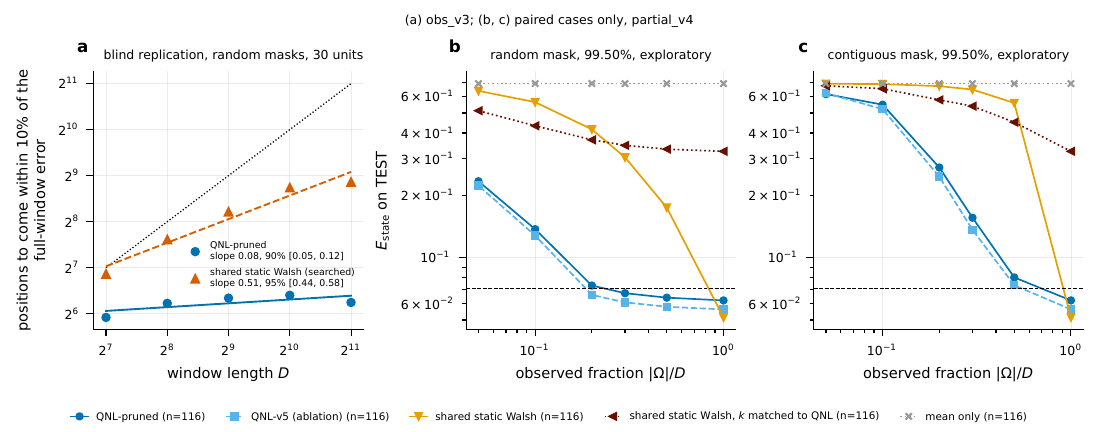}
  \caption{\textbf{With random observations, the number of positions QNL needs to approach its
  full-window accuracy is constant within a prespecified margin as the window grows, while the shared
  Walsh support's grows.}
  \textbf{a}, Preregistered blind replication (Supplementary Note~5): the number of randomly observed
  positions each method needs for its error to come within \ObsVthreeDelta\% of its own error with
  the whole window, against the window length $D$, both axes logarithmic; selected on VALIDATION from
  recordings never used as TEST before, central target fidelity (99.50\%). Points are geometric means
  over the cases where both methods deploy; lines are the mean per-unit slopes, given in the legend
  with the 90\% interval for QNL-pruned (the equivalence test, margin
  $\pm\ObsVthreeEquivalenceMargin$) and the 95\% interval for the Walsh support. The dotted line is
  $m=D$, the whole window. This is not the number of positions needed to meet $\epsilon$.
  \textbf{b}, \textbf{c}, Exploratory context, from the experiment that suggested the question:
  mean $E_{\mathrm{state}}$ on TEST against the observed fraction $|\Omega|/D$, random mask (b) and
  one contiguous gap (c), central target fidelity; the dashed line is $\epsilon$. Paired cases only:
  within each fraction every method is averaged over the cases where all of them deployed
  ($n=\PartialQnlNNinetyNineFiftyObsHundred{}$, in the legend); intervals are in Supplementary
  Table~5. For QNL, $c=b+A\hat z$ already determines the fixed circuit; the whole window is
  reconstructed only to score error. The matched-$k$ control is the shared static Walsh construction
  truncated to QNL's number of effective unknowns.}
  \label{fig:partial}
\end{figure}

How many observed positions each method needs as the window grows was tested in a preregistered
blind replication: protocol, hypotheses, tolerance and analysis were frozen before any of its data
existed, on new seeds and on recordings never used as TEST before, and the result was to be
reported whatever it showed. It replicates a design whose criterion had been chosen after a first
confirmatory test failed (Methods, Supplementary Note~5). The quantity is the number of randomly
observed positions a method needs for its error to come within \ObsVthreeDelta\% of its own error
with the whole window: the number needed to approach the full-window accuracy. QNL-pruned needed
\ObsVthreeRelGeoPrunedDOneTwoEight{} positions at $D=128$ and \ObsVthreeRelGeoPrunedDTwoZeroFourEight{}
at $D=2048$, from \ObsVthreeRelSharePrunedDOneTwoEight\% to \ObsVthreeRelSharePrunedDTwoZeroFourEight\%
of the window (Fig.~\ref{fig:partial}a), a slope on logarithmic axes of \ObsVthreeSlopePruned{}
(90\% interval \ObsVthreeSlopePrunedCI{}), equivalent to zero within the prespecified margin of
$\pm\ObsVthreeEquivalenceMargin$: equivalence, not a vanishing slope, since the interval excludes
zero. The searched
shared Walsh support needed \ObsVthreeRelGeoSearchedDOneTwoEight{} and
\ObsVthreeRelGeoSearchedDTwoZeroFourEight{}, a slope of \ObsVthreeSlopeSearched{}
(\ObsVthreeSlopeSearchedCI{}), and the paired difference of slopes was \ObsVthreeSlopeContrast{}
(95\% interval \ObsVthreeSlopeContrastCI{}). Both declared tests held, on the recordings never used
as TEST and on all recordings. On TEST, QNL-pruned's error with that many positions was
\ObsVthreeTestRatioPrunedMin{} to \ObsVthreeTestRatioPrunedMax{} times its full-window error, so the
tolerance held out of sample.

Three limits are part of the result. It holds for random observations only: with one contiguous gap
neither method saves positions, and both need nearly the whole window (slopes
\ObsVthreeContiguousSlopePruned{} and \ObsVthreeContiguousSlopeSearched{}). It concerns approaching
the full-window accuracy, not meeting the fidelity budget: QNL-pruned's full-window error already
lies close to $\epsilon$, and the number of positions needed to stay within $\epsilon$ grows with
$D$ (slope \ObsVthreeAbsSlopePruned{}, \ObsVthreeAbsSlopePrunedCI{}), so no claim is made about
it. And it depends on the tolerance: at 5\% instead of 10\% QNL-pruned's slope is
\ObsVthreeDeltaFiveSlopePruned{} (\ObsVthreeDeltaFiveSlopePrunedCI{}), outside the margin, though
still below the Walsh support's (Supplementary Note~5).

The question arose from an exploratory analysis of the deployed models of the main campaign, read
off TEST after the fact (Fig.~\ref{fig:partial}b,c; per dataset in Supplementary Figure~4): with the
whole window the shared static support is slightly more accurate, and with a random mask the order
reverses as soon as positions are hidden, while with a contiguous gap neither method saves
positions. Its aggregate over paired cases, a mean error within the budget down to three tenths of
the window, held for only \PartialQnlCasesMeetingNinetyNineFiftyObsThirty{} of
\PartialQnlNNinetyNineFiftyObsThirty{} cases individually and not at $D=128$; it is reported in full, over paired cases (Supplementary Table~5; unpaired in Supplementary Table~6) and with the per-length minima, in Supplementary Notes~4 and~5.

The shared support has \PartialWalshStaticUnknownsNinetyNineFiftyObsHundred{} effective unknowns
per window at the central fidelity, against \PartialQnlUnknownsNinetyNineFiftyObsHundred{} for
QNL-pruned, which raises the question of whether the advantage is simply having fewer unknowns.
A control answers it: the same shared Walsh construction truncated to as many coefficients as QNL
has effective unknowns, \PartialWalshMatchedKUnknownsNinetyNineFiftyObsHundred{}, reaches an error
of \PartialWalshMatchedKNinetyNineFiftyObsHundred{} with the whole window observed and stays above
QNL-pruned at every observed fraction. Having few unknowns is therefore not sufficient; the
advantage comes from which subspace those unknowns span. Plotted against observations per effective
unknown, the QNL and full shared-support curves overlap where their ranges meet, while the
matched-$k$ control stays well above both (Supplementary Note~4). The advantage is also not one of
numerical conditioning: measured over the directions the estimator actually uses, QNL's observed
design is the worse conditioned of the two (Supplementary Table~5).

In this task pruning is a cost. With a random mask at the central fidelity, QNL-v5, before
pruning, is more accurate than QNL-pruned at every observed fraction, with
\PartialQnlVfiveNinetyNineFiftyObsHundred{} against \PartialQnlNinetyNineFiftyObsHundred{} with the
whole window and \PartialQnlVfiveNinetyNineFiftyObsTwenty{} against
\PartialQnlNinetyNineFiftyObsTwenty{} with a fifth observed: the masks removed to save gates also
carried part of the subspace. The number of positions it needs to approach its own full-window
accuracy behaves as QNL-pruned's (slope \ObsVthreeVfiveSlope{}).

\subsection{Where the method stops}
\label{sec:limits}

The validation contract limits deployment to cases for which the selected model's one-sided error
bound and CX count meet the prespecified criteria; it is not a per-window certificate on TEST. A
method that abstains still leaves the state to be loaded, so coverage is a result in its own right.
At the central fidelity QNL-pruned deployed on
\CoveragePrunedNinetyNineFifty\% of the \MainCases{} cases, and on
\CoveragePrunedNinetyNineZero\%, \CoveragePrunedNinetyNineTwentyFive\% and
\CoveragePrunedNinetyNineSeventyFive\% at 99.00\%, 99.25\% and 99.75\%. The searched shared
Walsh support covers more at every fidelity: \CoverageSearchedNinetyNineFifty\% at the central
one, and \CoverageSearchedNinetyNineZero\%, \CoverageSearchedNinetyNineTwentyFive\% and
\CoverageSearchedNinetyNineSeventyFive\% at the other three in the same order
(Fig.~\ref{fig:limits}a, Supplementary Table~7; per dataset in Supplementary Figure~5). Its abstentions are
almost all failures of accuracy rather than of confidence or execution, and they are concentrated:
QNL-pruned deploys on none of the tested PTB-XL cases at any fidelity, where the baseline deploys
on part of them, and on none of the Paderborn cases at the strictest target. These abstentions
delimit the cases supported by the fitted affine family under the frozen protocol. The
per-dataset counts behind these statements are in Supplementary Table~7.

Because an abstaining case still has to be loaded somehow, we priced every case under two declared
fallback policies, without preferring either: an abstention is paid either at the exact diagonal's
cost, or at the searched Walsh support's cost with the exact diagonal behind it
(Fig.~\ref{fig:limits}b, Supplementary Table~8). The expected cost is the arithmetic mean of the
CX paid per case over all \MainCases{} cases, because it is an expectation: what a user pays on
average, abstentions included. The two policies answer differently. Under the exact fallback,
QNL-pruned costs \ExpectedPrunedExactNinetyNineFifty{} CX per case at the central fidelity against
\ExpectedSearchedExactNinetyNineFifty{} for the baseline, and it is the more expensive of the two at
every fidelity, from \ExpectedPrunedExactNinetyNineZero{} against
\ExpectedSearchedExactNinetyNineZero{} at 99.00\% to \ExpectedPrunedExactNinetyNineSeventyFive{}
against \ExpectedSearchedExactNinetyNineSeventyFive{} at 99.75\%: each of its abstentions is paid
at the exact diagonal's price, and it abstains more. Under the searched fallback the order
reverses by a small margin, \ExpectedPrunedSearchedNinetyNineFifty{} against
\ExpectedSearchedSearchedNinetyNineFifty{} at the central fidelity, and QNL-pruned is marginally
the cheaper at every fidelity. Which policy applies is a property of the deployment, not of the
method, and we report both.

A negative control tested whether any method deploys where there is no structure to learn. Each
window was permuted in place, which keeps its values, scale and phase range exactly and destroys
its temporal order, and every method was run on these surrogates for all seven views at $D=256$
and all four target fidelities (Fig.~\ref{fig:limits}c, Supplementary Table~9). QNL-pruned deployed
on \SurrogatePrunedDeployed{} of \SurrogatePrunedTotal{} surrogate runs, as did every other learned
shared model, while the per-sample Walsh support with an adaptive $k$ deployed on
\SurrogateAdaptiveDeployed{} of \SurrogateAdaptiveTotal{}, as did the exact diagonal, by
construction. The adaptive loader passes the contract by fitting each surrogate window on its own,
which is the behaviour a shared model cannot have. Restricted to the windows that do not already
meet $\epsilon$ with nothing loaded, the records support a lower bound on these counts rather than
an exact figure, which Fig.~\ref{fig:limits}c shows.
The result shows that the shared learned models do not pass the deployment contract after this
particular destruction of temporal structure. It does not constitute a universal
out-of-distribution detector: the contract is evaluated at case level on VALIDATION, and other
distribution shifts were not tested.

Two further checks bound the coverage result. Extending QNL's latent grid to $r=32$ and $r=64$,
on every case that had abstained for accuracy, rescued \AblationRescued{} of \AblationTotal{}
cases, all of them Paderborn views and nearly all at the strictest target; no PTB-XL case was
rescued (Fig.~\ref{fig:limits}d, Supplementary Table~10), a result consistent with a limitation of
the tested affine family rather than proof that latent dimension is irrelevant. This ablation is
reported on its own and
not folded into the coverage above. Finally, when TEST was made of units never seen in FIT,
QNL-pruned deployed on \UnitStressCoveragePruned\% of the \UnitStressRuns{} unit-split runs at
$D=512$ and the central fidelity, against \UnitStressRecordingCoveragePruned\% of the
\UnitStressRecordingRuns{} recording-split runs on the same views, with no detectable loss of
coverage (Supplementary Table~11).

\begin{figure}[tbp]
  \centering
  \includegraphics[width=\textwidth]{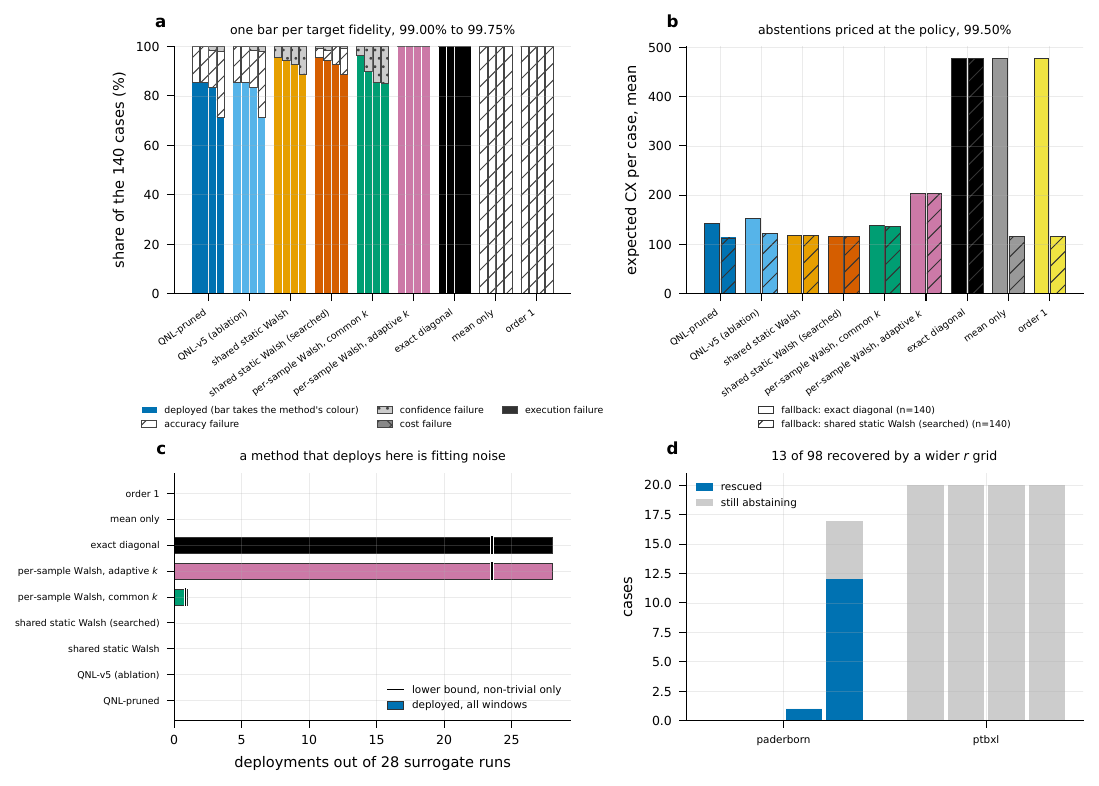}
  \caption{\textbf{Where the method stops working.}
  \textbf{a}, Coverage of each method over the \MainCases{} cases of the main campaign, one bar
  per target fidelity from 99.00\% to 99.75\%. The coloured part is the share deployed; the
  hatched parts above it are abstentions by declared cause.
  \textbf{b}, Expected two-qubit gate count per case at the central target fidelity (99.50\%),
  with each abstention priced at the declared fallback policy: exact diagonal (plain) or searched
  shared Walsh support backed by the exact diagonal (hatched). Bars are arithmetic means over the
  \MainCases{} cases, since an expected cost is an expectation; $n$ is given in the legend.
  \textbf{c}, Negative control: deployments on \NegativeControlRuns{} surrogate runs (seven views at
  $D=256$, four target fidelities) whose windows were permuted in place. The black tick is the
  lower bound on the count restricted to non-trivial windows; the per-window TEST errors needed to
  compute it exactly are not in the records.
  \textbf{d}, Extended latent grid ($r=32,64$) on the cases that abstained for accuracy, by
  dataset and target fidelity: \AblationRescued{} of \AblationTotal{} rescued. Reported as the
  ablation's own result and never folded into the coverage in \textbf{a}.}
  \label{fig:limits}
\end{figure}

\section{Discussion}
\label{sec:discussion}

The central empirical result is that, with random observations, the number of positions QNL needs
to approach its full-window accuracy is constant within a prespecified margin as the window grows,
whereas the strongest shared Walsh baseline needs more positions for longer windows; it comes from a
preregistered blind replication of a design whose criterion was chosen after a first confirmatory
test failed. It rests on a low-dimensional family embedded in a physically compiled Walsh circuit:
where both deploy, QNL-pruned and that baseline reach paired CX parity, and each window supplies
\CompressionNinetyNineFifty{} times fewer numbers to QNL. QNL therefore demonstrates not a cheaper
circuit but a lower-dimensional interface to a circuit of comparable cost. The difference lies in
what each window supplies, not in what the circuit receives: both circuits take one coefficient per
support term (\PairedKQnlNinetyNineFifty{} against \PairedKSearchedNinetyNineFifty{} on average over
the same pairs), so no reduction in the data transferred to the hardware is implied.

That interface follows from a learned physical family rather than from compression alone. Offline,
exact FIT Walsh coefficients determine candidate supports; the affine map $c(z)=b+Az$ is fitted on
FIT, supports are searched and pruned using transpiled CX, and VALIDATION selects under the
deployment contract. The selected $S$, $b$ and $A$ then remain fixed. Online, a compatible window
is projected onto $m+\operatorname{span}(B)$ by a linear least-squares solve, after which
$c=b+A\hat z$ gives the $K$ physical coefficients. Thus the $r$ latent coordinates are not
themselves $K$ gate angles. They are a lower-dimensional set of unknowns from which the parameters
of one fixed topology are evaluated, with no per-input circuit optimization or model training.

Partial observation is the operational consequence of this interface, and its scaling has a direct
mechanism. From positions $\Omega$, QNL solves only for its $r$ coordinates, which already determine
the circuit coefficients. The number of unknowns is $r$ whatever the window length: the deployed
models have $r=\ObsVthreeMeanLatent$ on average, and in the replication QNL-pruned needed between
\ObsVthreePerCoordinateMin{} and \ObsVthreePerCoordinateMax{} observed positions per latent coordinate
to approach its full-window accuracy, at every $D$ from 128 to 2048. A longer window adds positions
but not coordinates, so the number needed stays constant within the prespecified margin and the
fraction of the window to observe falls. A shared Walsh support estimates its coefficients directly
and needs more observations as the window grows, and a Walsh support with as many coefficients as
QNL has coordinates does not close the gap: the advantage comes from which subspace the unknowns
span, not only from how few they are. The limits stated with the result apply here unchanged: random
observations only, approaching the full-window accuracy rather than meeting the budget, and
dependence on the tolerance. The first confirmatory test failed precisely because it measured the
budget, which a deployed model meets only narrowly.

Estimating coordinates in a fixed subspace from random entries is governed by its coherence with
the sampling basis and scales with $r$, not
$D$\cite{donoho2006compressed,candes2007incoherence,candes2009matrix,balzano2010subspaces}: the
Walsh constraint supplies that coherence, since $\mathrm{span}(B)\subseteq\mathrm{span}(W_S)$
bounds $\mu\leq K/r$ for every $D$, and a contiguous gap is the coherent extreme. That coherence
is unmeasured here, so this is a mechanism, not a bound.

The result is distributional rather than universal. Complete FIT examples are required to learn
the family. After that family has been frozen, a new signal compatible with it can be instantiated
from partial classical observations under the tested mask and least-squares protocol. This does
not imply sublinear-time access to an arbitrary vector, and it does not establish lower end-to-end
wall-clock runtime. It shows that, with random observations, QNL needs a number of classical positions that is constant within a prespecified margin as the window grows, to approach its full-window accuracy, and that it produces circuit parameters directly from the inferred latent coordinates, without a complete classical reconstruction as an intermediate deployment object. Whether reading fewer positions lowers the total cost depends on the condition derived in Supplementary Note~3, which compares the cost of the positions saved with the differences in inference, parameter handling, execution and amortised learning; none of these terms was measured.

This positioning distinguishes QNL from two neighbouring bodies of work. Target-specific Walsh,
Fourier, function-loading, tensor-network and compilation methods determine coefficients,
factorizations or circuits from the complete target supplied at deployment
\cite{welch2014diagonal,zylberman2024walsh,moosa2024fsl,rosenkranz2025multivariate,dilip2022compression,jumade2023shortdepth,jobst2024mps,alonsolinaje2025compilation,zhang2026aqer}.
Learned approaches already amortize state preparation: transformer models can generate circuits
for unseen amplitude vectors, neural encoders can predict the parameters of a fixed ansatz from
complete images, and adaptive transforms can learn a shared
basis\cite{daimon2024transformer,aoun2026neural,budiutama2026aiqt}. QNL's contribution is not any
one of these ingredients in isolation. It is their specific combination in a phase-loading
setting: a low-dimensional learned family, latent coordinates as the online interface, an affine
map to one fixed physical Walsh support, CX-aware support selection and pruning, a statistical
deployment contract, and latent inference from incomplete observations.

The deployment contract makes the scope visible but does not eliminate failure. QNL-pruned covers
fewer cases than the searched baseline and none of the tested PTB-XL cases, and extending the latent
grid to $r=64$ rescues Paderborn cases but no PTB-XL case. This is consistent with the tested affine
family being inadequate for those electrocardiogram windows; it does not show that
electrocardiograms in general, or a nonlinear learned family, cannot be loaded compactly. Likewise,
deployment on \SurrogatePrunedDeployed{} of \SurrogatePrunedTotal{} temporally permuted surrogate
runs supports the intended conservative behaviour under that negative control, but it is not a
general guarantee of out-of-distribution detection.

The resource claims should remain separated. Quantum circuit cost is measured by transpiled CX
and depth; per-instance interface dimension by $r$ for QNL and $k$ for the shared Walsh baselines;
classical acquisition by $|\Omega|/D$ under masked held-out windows, as a count of positions read
and not as a cost (Supplementary Note~3). Online inference is a least-squares solve followed by
$b+A\hat z$, and offline learning includes the FIT initialization, transforms, support search,
decoder fits, transpilation-aware pricing, pruning and validation selection; neither was benchmarked
end to end. The expected gate count per case depends on the declared fallback, favouring the baseline
under the exact one and QNL marginally under the searched one, and is a count under a policy, not a
runtime.

Several limitations define the next experiments. No circuit was executed on hardware, and the
noise analysis is an analytic estimate from gate counts. The largest windows use eleven qubits, and
the scaling of the number of positions was tested from 128 to 2048 samples only: whether it stays
constant within the prespecified margin beyond that range is a prediction of the mechanism above, not a result.
The partial-observation experiments mask already acquired held-out data, and the seeds of the
replication re-split the same recordings rather than drawing new ones; a prospective study with a
physical acquisition system, new recordings and longer windows is the direct test. Structured
sampling designs that would bring the advantage to contiguous gaps are untested. A complete
end-to-end comparison must time acquisition, online inference, parameter transfer, circuit execution
and fallback, and is not supplied here. Nonlinear decoders are a plausible route beyond the affine
failures seen on PTB-XL, but their coverage and physical cost remain future work. Hardware tests,
prospective partial acquisition and an end-to-end resource study are therefore required before
making claims about runtime or application-level advantage.

\section{Methods}
\label{sec:methods}

\subsection{Phase encoding and error metric}
\label{sec:encoding}

A window $x\in\mathbb{R}^{D}$, $D=2^{n}$, is loaded as the diagonal phase state
$\ket{\psi_x}=D^{-1/2}\sum_{i}e^{i\lambda x_i}\ket{i}$, prepared by a diagonal unitary acting on
the uniform superposition $H^{\otimes n}\ket{0}$. A loader produces a phase vector $g$ in place of
$x$, and its error on that window is
\begin{equation}
  E_{\mathrm{state}}=\sqrt{1-F},\qquad
  F=\Bigl|\tfrac{1}{D}\textstyle\sum_{i}e^{i\lambda(g_i-x_i)}\Bigr|^{2},
\end{equation}
the trace distance between the loaded and the exact state. It bounds the change of every
expectation value, $|\langle O\rangle_g-\langle O\rangle_x|\le 2\lVert O\rVert E_{\mathrm{state}}$,
and it is invariant to a global phase. Because every loader considered here is diagonal, $F$ is
computed in closed form from the two phase vectors, without simulating a circuit and without
sampling; $1-F$ is assembled from half-angle sines so that it does not cancel to rounding error for
accurate models.

The four target fidelities are \TargetFidelities{}, with the error budget derived from each as
$\epsilon=\sqrt{1-F}$; where one fidelity is needed it is the central one, \CentralFidelity{}.
Circuits are built in Qiskit\cite{javadiabhari2024qiskit} and transpiled to the basis
\BasisGates{} at optimisation level \OptimizationLevel{} with transpiler seed \TranspileSeed{}.
CX is the number of two-qubit gates in the transpiled circuit and depth its transpiled depth; no
circuit uses ancillas.

\subsection{The learned family}
\label{sec:family}

QNL maps $r$ sample-specific coordinates $z$ to the coefficients of a shared Walsh support $S$
through an affine decoder,
\begin{equation}
  c(z)=b+Az,\qquad g(z)=W_{S}\,c(z)=m+Bz,\quad m=W_{S}b,\quad B=W_{S}A,
\end{equation}
where the columns of $W_{S}$ are the Walsh functions indexed by the masks in $S$. Each Walsh term is
a phase gadget whose angle is one coefficient, so the $K=|S|$ gadgets form one circuit whose
topology is fixed for the whole dataset, and a window changes only its angles. Three quantities are
kept distinct throughout: $r$, the numbers a window contributes; $K$, the Walsh terms the circuit
realises; and CX, the two-qubit gates of that circuit after transpilation. Every signal the
deployed model can load lies in the affine subspace $m+\operatorname{span}(B)$, of dimension at
most $r$.

The model is fitted on FIT alone, in two stages, after the Walsh coefficients of the complete FIT
windows have been computed exactly; this transform is neither an approximation nor a circuit
optimization. First, a non-linear generative family, one of the
\emph{template}, \emph{harmonic}, \emph{bank} and \emph{gadget} architectures, is trained jointly
over the FIT windows for each $r$ in \RGrid{}, with its linear parameters profiled out by variable
projection\cite{golub1973varpro}; the architecture and $r$ are selected on VALIDATION. This family is
used only to seed the FIT coordinates $z$ and to propose supports: it is never deployed and never
evaluated on TEST. Second, the affine decoder is fitted to the Walsh coefficients of the FIT windows
restricted to $S$ by alternating least squares, starting from the seeded coordinates: with $z$
fixed, $(b,A)$ is the least-squares solution with a ridge of \DecoderRidge{} on $A$ and none on the
intercept $b$; with $(b,A)$ fixed, each window's $z$ is the least-squares solution; the alternation
runs for \DecoderRounds{} rounds. The model that is deployed is therefore affine. Once a model is selected, $S$, $b$ and $A$---equivalently $m=W_Sb$ and
$B=W_SA$---and the circuit topology are frozen. Online, each TEST window requires only a
least-squares projection onto this fixed subspace followed by evaluation of $c=b+A\hat z$: no support
search, decoder refit, per-input circuit optimization or model training is performed on TEST.

The coordinates of a new window are its least-squares projection onto the model's subspace,
$\hat z=\arg\min_{z}\lVert x-m-Bz\rVert_2$, computed with \texttt{numpy.linalg.lstsq} at its default
tolerance on the whole window in the main campaigns, and restricted to the observed positions,
with a declared tolerance, under partial observation (see below; Supplementary Note~6). Because the Walsh columns are
orthogonal, the projection in signal space and the least-squares step in coefficient space used
during fitting have the same minimiser. Least squares on the phase is a surrogate for
$E_{\mathrm{state}}$, not $E_{\mathrm{state}}$ itself: it minimises the Euclidean distance between
phase vectors, which approximates the trace distance when the phase error is small.

Earlier versions of the project also compiled signals through quantum signal processing and its
generalisations\cite{low2017qsp,gilyen2019qsvt,motlagh2024gqsp} and through the approximate
quantum Fourier transform\cite{coppersmith1994aqft,barenco1996aqft}. None of these constructions was
selected, and the final backend compiles Walsh phase gadgets only.

\subsection{Support search and backward pruning}
\label{sec:support}

Each Walsh term $e^{i\theta Z_{s}}$ on the qubits of mask $s$ is compiled as a phase
gadget\cite{cowtan2020gadgets}. Commuting gadgets are emitted either as individual CNOT ladders,
in the order that minimises the transpiled CX among a Gray-like order, a nearest-neighbour order,
their two-opt improvements and, for small supports, the exact optimum, or as a single parity
network synthesised by GraySynth\cite{amy2018graysynth}; both are transpiled and the one with fewer
CX is kept. Every support that enters a decision is priced this way, with the transpiled circuit
and not with a proxy, and each distinct support is compiled once.

QNL-v5 ranks the Walsh masks by the energy of the FIT coefficients that the seeded coordinates
predict, divided by a local gate cost, fits a decoder on each prefix of that ranking over a fixed
grid of support sizes, and keeps the candidates that meet the deployment contract on VALIDATION.
From the cheapest of them, a parity-aware local search evaluates at most \MaxSupportProposals{}
further supports and accepts a move only if the contract still holds at the target fidelity.

QNL-pruned starts from the support QNL-v5 selects and may only remove masks. A ladder of target
sizes descends from $K_{\mathrm{v5}}$ in steps of \PruneStep{} down to a fraction \PruneMinFraction{}
of it. At each step the masks are shortlisted by the FIT energy the model would lose by dropping
them, which the fitted design determines exactly because each column of the least-squares
objective is fitted independently: $R(S)=\lVert C\rVert^{2}-\sum_{s\in S}\lVert P_{D}c_{s}\rVert^{2}$,
with $C$ the FIT coefficients and $P_{D}$ the projector onto the columns of $[1\,|\,Z]$. The
\PruneShortlist{} masks with the smallest loss are each removed in turn, the decoder is refitted
for real after every removal, and the removal that costs the least FIT error is kept. The search
reads FIT only. VALIDATION then applies the contract to the frozen candidates, QNL-v5 included, and
selects the one with the fewest two-qubit gates, then the smallest depth, then the smallest $K$.
Pruning never changes $r$, never admits a mask outside the QNL-v5 support and keeps a single
topology; the test suite checks each of these properties, that every deletion is refitted, that
the path is deterministic, and that QNL-v5 is always in the pool as the fallback.

\subsection{Baselines}
\label{sec:baselines}

Nine methods are evaluated on every case, all as diagonal phase loaders built from Walsh gadgets
and priced by the same compiler. \emph{Mean only} loads the FIT template $\mu$ and nothing per
window. \emph{Order 1} adds the $n$ single-qubit Walsh masks, whose gadgets are single $R_z$
rotations, so it spends $\log_2 D$ numbers per window at no two-qubit cost beyond $\mu$. The
\emph{shared static Walsh} support keeps the $k$ masks of largest mean FIT energy, the same for
every window, with each window's $k$ coefficients obtained by projection\cite{welch2014diagonal,zylberman2024walsh}; $k$ is chosen on VALIDATION as the cheapest size that meets the contract. The
\emph{searched} variant gives that baseline QNL's own support search, with the same size grid, the
same contract, the same proposal budget and the same compiler, so that what separates it from QNL
is the latent model alone; it is the strongest physical baseline. The \emph{per-sample} Walsh
supports keep each window's own largest $k$ coefficients, either with one $k$ for every window,
chosen on VALIDATION, or with the smallest $k$ that meets $\epsilon$ on each window. Because a
per-sample loader must be able to load any window, its reported CX is the maximum over the windows
it loads. The \emph{exact} diagonal realises all $D-1$ non-constant Walsh gadgets with one
parameter each\cite{bullock2004diagonal}, has no representation error and is transpiled once.

These nine do not include the learned loaders discussed in the Introduction, and the omission
follows from the quantity being measured rather than from their performance. A transformer decoder
that emits a gate sequence, a neural encoder that maps a complete image to the parameters of a
fixed topology, and an adaptive interpolating transform that performs a full classical transform
and then selects top-$k$ coefficients all consume the complete instance at
deployment\cite{daimon2024transformer,aoun2026neural,budiutama2026aiqt}. Their per-instance
classical interface therefore has dimension $D$ by construction, so neither the comparison at
equal numbers per window nor the partial-observation experiment is defined for them: there is no
smaller set of per-instance unknowns to match or to estimate. They could be compared on circuit
cost alone, and for that purpose the searched shared Walsh support is the more informative test,
because it differs from QNL in the latent model and in nothing else.

In the comparison at equal numbers per window, each Walsh baseline is matched to QNL with $k=r$.
The adaptive per-sample support is excluded from it: with the count used as a cap and no budget to
adapt to, every window spends the whole cap, and its row is identical to that of the common-$k$
support by construction.

\subsection{Datasets and preprocessing}
\label{sec:data}

Seven views of five public datasets were used: the current and voltage channels of
PLAID\cite{medico2020plaid} and WHITED\cite{kahl2016whited}, the motor current of the Paderborn
bearing dataset\cite{lessmeier2016paderborn}, the current channel of HIFDA\cite{navarro2025hifda},
and the electrocardiograms of PTB-XL\cite{wagner2020ptbxl}, distributed through
PhysioNet\cite{goldberger2000physionet}. Each recording is cut into consecutive, non-overlapping
native windows of \NativeLength{} samples, \NativeLengthPtbxl{} for PTB-XL, never spanning two
recordings. A view of dimension $D$, one of \ViewDimensions{}, keeps every $(\text{native}/D)$-th
sample with no anti-aliasing filter, so that each view is an exact subsample of the next and
dimensions are comparable; the energy that this folds above the Nyquist frequency of the view is
reported per dataset and dimension (Supplementary Table~12); sampling rates, units, recordings and windows per dataset are in Supplementary Table~13.

The scale $\sigma$ and the phase factor $\lambda$ are computed on the native FIT windows, with the
mean of each window removed, and shared across every $D$: $\sigma$ is the pooled standard deviation,
and $\lambda=\PhaseMargin\,\pi/q$, with $q$ the \PhaseQuantile{} quantile of $|x|/\sigma$ over FIT.
The FIT mean window $\mu$ is a fixed template that is subtracted before learning and paid for in CX
by every loader that reproduces it. The main split is \SplitShares{} over whole recordings, seeded,
so TEST contains new recordings of units seen in FIT; PTB-XL uses the patient-separated folds its authors recommend, assigned in order to FIT, VALIDATION and TEST. A window is \emph{trivial} at a target fidelity
if it already meets $\epsilon$ with nothing loaded, that is with the uniform state.

The voltage channel of HIFDA was excluded. Its acquisition band is 300\,Hz to 50\,kHz, as the
dataset's documentation states\cite{navarro2025hifda}, so the 50\,Hz mains fundamental is suppressed: \HifdaVoltageFundamentalShare\% of a window's energy lies in the bin of the fundamental and \HifdaVoltageEnergyAboveOneKilohertz\% above 1\,kHz. Subsampled without a filter, its views lose most of that energy above their Nyquist frequency, from \HifdaVoltageAliasingMin\% at the largest view to \HifdaVoltageAliasingMax\% at the smallest, so they record aliased energy rather than the signal. Its scale is normalised by $\lambda$ like every other view, so the exclusion rests on the aliasing and not on the channel's amplitude. Its measurements are reported with those of the included views, marked as
excluded (Supplementary Table~12).

\subsection{Deployment contract and abstention}
\label{sec:contract}

A method is deployed on a case at a target fidelity if and only if
\begin{equation}
  \mathrm{UCL}_{\mathrm{VAL}}(E_{\mathrm{state}})\le\epsilon
  \quad\text{and}\quad \mathrm{CX}<\mathrm{CX}_{\mathrm{exact}},
\end{equation}
where $\mathrm{UCL}$ is the one-sided \ContractConfidence{} upper confidence limit of the mean
VALIDATION error, clustered by recording. With $G$ recordings, group sums $S_g$ of $n_g$ windows,
$N$ windows in all and mean error $\bar e$, the cluster-robust standard error is
$\sqrt{G/(G-1)\sum_{g}(S_{g}-n_{g}\bar e)^{2}}/N$\cite{cameron2008cluster}, and the limit adds to
$\bar e$ that error times the Student quantile with $G-1$ degrees of freedom, so windows of one
recording count once. The cost half of the contract applies to every method alike. A method that
is not deployed abstains, and the abstention is classified as a failure of accuracy when the mean
VALIDATION error exceeds $\epsilon$, of confidence when the mean meets $\epsilon$ but its upper limit
does not, of cost when the accuracy contract holds but the circuit is not cheaper than the exact
one, and of execution when the case could not be run. Every planned case is reconciled against the
canonical registry, so that a missing record becomes an execution failure instead of shrinking the
totals.

TEST is materialised once per case, after every selection has been frozen, and never enters a
choice. On TEST each deployed model reports its mean error, its upper confidence limit, the
achieved fidelity $1-E^{2}$ averaged over windows, CX, depth and $K$. Success is read twice: over
all TEST windows, and restricted to the non-trivial ones, the second as a sensitivity analysis.

\subsection{Comparisons and statistics}
\label{sec:statistics}

Two comparisons are made. At equal numbers per window, methods are compared by their TEST error
with $r$ or $k$ in \RGrid{} and no target fidelity imposed. At a target fidelity, each method
chooses its own size under the contract and methods are compared by the resources they need.
Ratios between methods are paired within a case and a fidelity, over the cases where both deploy,
and summarised by their geometric mean with a 95\% Student interval taken on the log scale; no
interval is given with fewer than \MinReplicates{} replicates. A paired ratio inside
$[\BandLow,\BandHigh]$ is read as a tie. Aggregates are computed per target fidelity and never
pooled across fidelities.

The expected cost per case counts abstentions: a case costs the method's own CX where it deploys and
the fallback's CX where it does not, under two declared policies, fallback to the exact diagonal,
or to the searched shared Walsh support with the exact diagonal behind it. It is aggregated as an
arithmetic mean over cases. The frozen protocol specified a geometric mean; the arithmetic mean was
adopted after the results were available, because an expected cost is an expectation and a
geometric mean shrinks precisely the large costs of abstentions paid at the exact diagonal's price.
Both aggregates are reported (Supplementary Table~8, Supplementary Note~7). The supports each method priced with a transpiled circuit and the candidates it judged on VALIDATION are counted in Supplementary Table~14.

The total fidelity under gate noise is an analytic estimate, not a simulation:
$F_{\mathrm{total}}(p)=(1-E_{\mathrm{test}}^{2})(1-p)^{\mathrm{CX}}$ for a two-qubit gate error $p$
in \GateErrors{}, assuming independent depolarising error per two-qubit gate and ignoring idling,
crosstalk and readout. The three methods it compares are evaluated over the same cases.

\subsection{Campaigns}
\label{sec:campaigns}

The main campaign covered every view at $D$ in \MainDimensions{}, \SeedCount{} seeds and the four target fidelities, \MainCases{} cases in all. The scaling campaign added $D=\ScalingDimension{}$ at
the central fidelity with seeds \ScalingSeeds{}, for every view whose native window allows it, and
the exponent of $\mathrm{CX}\propto D^{\alpha}$ is fitted by least squares on the logarithms of the
geometric-mean CX at each $D$. The unit-stress campaign ran at $D=\UnitStressDimension{}$ and the
central fidelity with TEST made of units never seen in FIT: datasets with at least
\UnitSplitMinUnits{} units were split \SplitShares{} over whole units, and the others by
leave-one-unit-out folds, \UnitFoldsPaderborn{} for Paderborn and \UnitFoldsHifda{} for HIFDA,
drawn from a permutation of the unit identifiers with seed \UnitFoldSeed{}. The negative control
ran every method at $D=\NegativeControlDimension{}$, seed \NegativeControlSeed{} and the four target
fidelities on surrogates that permute the samples within each window, which keeps each window's
values, and therefore its $\sigma$ and its phase range, exactly, and destroys its temporal order; a
test checks both properties against the real views. Its primary reading is restricted to the
non-trivial windows. The $r$ ablation extended the latent grid to \RGridAblation{} on every case
whose QNL-pruned record in the main campaign abstained for accuracy, all of them, selected by that
rule alone and with no case added or removed afterwards; it is reported separately and never folded
into the coverage of the main campaign.

\subsection{Partial observation}
\label{sec:partial-methods}

The deployed models of the main campaign were reused without refitting (Supplementary Note~4; results per dataset in Supplementary Table~15). Because the campaign
records metrics and not models, each model was recomputed from its frozen source at its target
fidelity, and the case was analysed only if the recomputed $K$, CX and VALIDATION error matched the recorded ones. For each TEST window, the positions in a mask $\Omega$ were kept and the rest
discarded, with observed fractions \ObservedFractions{}, under two patterns drawn from the case's
seed: positions sampled uniformly without replacement, and one contiguous run with a random start.
The unknowns were estimated by
\begin{equation}
  \hat u=\arg\min_{u}\lVert M_{\Omega}u-(x_{\Omega}-m_{\Omega})\rVert_{2},
\end{equation}
the minimum-norm least-squares solution, with the same call and the same truncation tolerance,
\PartialRcond{} relative to the largest singular value, for every method; only the design $M$
changes, $B$ for QNL, $W_{S}$ for a shared Walsh support and none for the template. For QNL, these
estimated coordinates determine the deployed circuit directly through $c=b+A\hat z$; no complete
classical window is required to evaluate its angles. The window was nevertheless reconstructed at every position solely to measure its error against the held-out complete window, as in the recovery of structured signals from incomplete measurements\cite{donoho2006compressed,candes2009matrix}; no support, decoder or circuit was refitted or selected per window. The per-sample
baselines were excluded, since their support is chosen from the very window that is not observed.
A control used the shared static Walsh construction truncated to as many coefficients as QNL's
effective rank on the same case. Comparisons are paired: within each fidelity, pattern and fraction,
every method is averaged over the cases on which all of them deployed. Conditioning is reported as
$\sigma_{\max}/\sigma_{\mathrm{kept}}$ over the singular directions the estimator retains at the
tolerance, beside $\sigma_{\max}/\sigma_{\min}$ over all of them, both aggregated per cell as
geometric means.

\subsection{Observation scaling}
\label{sec:obs-methods}

How many observed positions each method needs as the window grows was tested in a preregistered
blind replication, under its own frozen protocol (version~3 of PROTOCOL\_obs\_scaling; versions~1
and~2, its antecedents, are described with their outcomes in Supplementary Note~5). Its protocol,
hypotheses, tolerance and analysis were committed with their SHA256 before any of its data existed,
and it was declared to be reported whatever it showed. For seeds \ObsVthreeSeedFirst{} to
\ObsVthreeSeedLast{}, new state preparations were learned by the same pipeline as the main campaign
and deployed by the unmodified evaluation code at the central target fidelity, for $D$ from
\ObsVthreeDimensionFirst{} to \ObsVthreeDimensionLast{}; before any run, every file of that code was
checked against the commit the main evaluation ran from, and each deployed model was recomputed and
verified against its record as above. Observations followed an absolute axis of
\ObsVthreeAxisPoints{} counts from \ObsVthreeAxisFirst{} to \ObsVthreeAxisLast{}, clipped to $D$, with
nested masks: each window has one fixed random order of its positions, and the mask of size $m$ is
its first $m$ positions, so that errors are comparable along $m$. The estimator is the one above.

For each method, case and $D$, the selected count $m_{\mathrm{rel}}$ is the smallest $m$ whose mean
VALIDATION error is at most $1+\delta$ times the same method's mean VALIDATION error with the whole
window, with $\delta=\ObsVthreeDelta\%$. It always exists, and no margin below $\epsilon$ can move
it. The primary analysis uses only VALIDATION and TEST windows from recordings that were never on
the TEST side of any partition for seeds 0 to \ObsVthreeOriginSeedLast{}, which are those of the
earlier versions and of the exploration. TEST was read only at the selected counts and with the whole
window. Per unit, a (dataset, seed) with at least \ObsVthreeMinDimensionsPerUnit{} window lengths at
which both QNL-pruned and the searched shared Walsh support deployed, the slope of
$\log m_{\mathrm{rel}}$ on $\log D$ was fitted by least squares, and slopes were averaged over units.
Two tests were declared. H1: QNL-pruned's mean slope is equivalent to zero, its 90\% Student
interval inside $\pm\ObsVthreeEquivalenceMargin$ (two one-sided tests at $\alpha=0.05$). H2: the
paired difference of slopes, Walsh support minus QNL-pruned, has its 95\% Student interval above
zero. The statement that the count is constant within a prespecified margin is made only if both
hold. The protocol also declared the same analysis on all recordings, at $\delta$ of 5\% and 20\%,
with the absolute rule of version~1 (the smallest $m$ whose VALIDATION upper limit is at most
$\epsilon$) and with contiguous masks, all reported in Supplementary Note~5.

\subsection{Reproducibility}
\label{sec:reproducibility}

Each protocol is frozen and hashed (Supplementary Note~1 gives every version), and every run records the SHA256 of the protocols that govern
it (\texttt{\FinalProtocolSha} for the evaluation, \texttt{\DataSpecSha} for the data specification,
\texttt{\PartialProtocolSha} for partial observation), the hashes of the source files it ran
under, the git commit and whether the working tree was clean. Every runner reconciles its records
against the canonical case registry.

Three checks measure reproducibility. Of the \DeployedPrunedModels{} QNL-pruned models deployed in
the main campaign, \ReproducedPrunedModels{} were recomputed from their frozen sources and matched
the recorded $K$, CX and VALIDATION error. The models that were not reproduced are the four target fidelities of a single case,
refused as a whole because its recomputed seeding model differs from the recorded one only in one
fitted frequency, beyond the relative tolerance of \BlockTolerance{} used to compare them. The
only other loss is not among these models: one cell of the scaling campaign, never deployed,
because the support search of its source run does not land on the same support when repeated. Two executions of the partial-observation campaign, before and after a correction to
its conditioning diagnostics, agree bit for bit in \BitIdenticalCells{} of \ComparedCells{} cells in
every field other than conditioning. Finally, and unrelated to those losses, \RankDeficientModels{} of the \RankCheckedModels{}
models that were reproduced span fewer directions in signal space than their nominal $r$, all of
them PLAID views with the \emph{bank} architecture: they deploy and are analysed like any other,
with a subspace smaller than $r$ says. These checks are detailed in Supplementary Note~8, and the remaining deviations from the protocol and the measurements it asked for and did not obtain are listed in Supplementary Note~9.

\subsection*{Data availability}

The datasets analysed are public and available from their original
sources\cite{medico2020plaid,kahl2016whited,lessmeier2016paderborn,navarro2025hifda,wagner2020ptbxl,goldberger2000physionet};
they are not redistributed here. The prepared windows are regenerated from them by the script
included with the code. The records of every campaign, the frozen state preparations of seeds 0
to 4, and every table, figure and number of this article are archived at \ArchiveDOI{} and
available at \RepositoryURL{}.

\subsection*{Code availability}

The code, the frozen protocols with their version history and the scripts that regenerate every
table, figure and number of this article from the records are available at \RepositoryURL{} under
the MIT licence, and the version used for this article is archived at \ArchiveDOI{}.

\bibliographystyle{unsrtnat}
\bibliography{references}

@article{aaronson2015fineprint,
  author  = {Aaronson, Scott},
  title   = {Read the fine print},
  journal = {Nature Physics},
  volume  = {11},
  pages   = {291--293},
  year    = {2015}
}

@article{giovannetti2008qram,
  author  = {Giovannetti, Vittorio and Lloyd, Seth and Maccone, Lorenzo},
  title   = {Quantum Random Access Memory},
  journal = {Physical Review Letters},
  volume  = {100},
  pages   = {160501},
  year    = {2008}
}

@misc{grover2002superpositions,
  author        = {Grover, Lov and Rudolph, Terry},
  title         = {Creating superpositions that correspond to efficiently integrable probability distributions},
  year          = {2002},
  eprint        = {quant-ph/0208112},
  archivePrefix = {arXiv}
}

@article{mottonen2005uniformly,
  author  = {M{\"o}tt{\"o}nen, Mikko and Vartiainen, Juha J. and Bergholm, Ville and Salomaa, Martti M.},
  title   = {Transformation of quantum states using uniformly controlled rotations},
  journal = {Quantum Information and Computation},
  volume  = {5},
  number  = {6},
  pages   = {467--473},
  year    = {2005}
}

@article{shende2006synthesis,
  author  = {Shende, Vivek V. and Bullock, Stephen S. and Markov, Igor L.},
  title   = {Synthesis of quantum-logic circuits},
  journal = {IEEE Transactions on Computer-Aided Design of Integrated Circuits and Systems},
  volume  = {25},
  number  = {6},
  pages   = {1000--1010},
  year    = {2006}
}

@article{plesch2011universal,
  author  = {Plesch, Martin and Brukner, {\v{C}}aslav},
  title   = {Quantum-state preparation with universal gate decompositions},
  journal = {Physical Review A},
  volume  = {83},
  pages   = {032302},
  year    = {2011}
}

@article{zhang2022optimaldepth,
  author  = {Zhang, Xiao-Ming and Li, Tongyang and Yuan, Xiao},
  title   = {Quantum State Preparation with Optimal Circuit Depth: Implementations and Applications},
  journal = {Physical Review Letters},
  volume  = {129},
  pages   = {230504},
  year    = {2022}
}

@article{sun2023asymptotically,
  author  = {Sun, Xiaoming and Tian, Guojing and Yang, Shuai and Yuan, Pei and Zhang, Shengyu},
  title   = {Asymptotically optimal circuit depth for quantum state preparation and general unitary synthesis},
  journal = {IEEE Transactions on Computer-Aided Design of Integrated Circuits and Systems},
  volume  = {42},
  year    = {2023}
}

@article{araujo2021divide,
  author  = {Araujo, Israel F. and Park, Daniel K. and Petruccione, Francesco and da Silva, Adenilton J.},
  title   = {A divide-and-conquer algorithm for quantum state preparation},
  journal = {Scientific Reports},
  volume  = {11},
  pages   = {6329},
  year    = {2021}
}

@article{bullock2004diagonal,
  author  = {Bullock, Stephen S. and Markov, Igor L.},
  title   = {Asymptotically optimal circuits for arbitrary n-qubit diagonal computations},
  journal = {Quantum Information and Computation},
  volume  = {4},
  number  = {1},
  pages   = {27--47},
  year    = {2004}
}

@article{welch2014diagonal,
  author  = {Welch, Jonathan and Greenbaum, Daniel and Mostame, Sarah and Aspuru-Guzik, Al{\'a}n},
  title   = {Efficient quantum circuits for diagonal unitaries without ancillas},
  journal = {New Journal of Physics},
  volume  = {16},
  pages   = {033040},
  year    = {2014}
}

@article{zylberman2024walsh,
  author  = {Zylberman, Julien and Debbasch, Fabrice},
  title   = {Efficient quantum state preparation with {W}alsh series},
  journal = {Physical Review A},
  volume  = {109},
  pages   = {042401},
  year    = {2024},
  doi     = {10.1103/PhysRevA.109.042401}
}

@article{moosa2024fsl,
  author  = {Moosa, Mudassir and Watts, Thomas W. and Chen, Yiyou and Sarma, Abhijat and McMahon, Peter L.},
  title   = {Linear-depth quantum circuits for loading {F}ourier approximations of arbitrary functions},
  journal = {Quantum Science and Technology},
  volume  = {9},
  pages   = {015002},
  year    = {2024},
  doi     = {10.1088/2058-9565/acfc62}
}

@inproceedings{holmes2020smooth,
  author    = {Holmes, Adam and Matsuura, A. Y.},
  title     = {Efficient quantum circuits for accurate state preparation of smooth, differentiable functions},
  booktitle = {2020 IEEE International Conference on Quantum Computing and Engineering (QCE)},
  pages     = {169--179},
  year      = {2020}
}

@article{marinsanchez2023approximate,
  author  = {Marin-Sanchez, Gabriel and Gonzalez-Conde, Javier and Sanz, Mikel},
  title   = {Quantum algorithms for approximate function loading},
  journal = {Physical Review Research},
  volume  = {5},
  pages   = {033114},
  year    = {2023}
}

@article{gonzalezconde2024polynomial,
  author  = {Gonzalez-Conde, Javier and Watts, Thomas W. and Rodriguez-Grasa, Pablo and Sanz, Mikel},
  title   = {Efficient quantum amplitude encoding of polynomial functions},
  journal = {Quantum},
  volume  = {8},
  pages   = {1297},
  year    = {2024}
}

@article{rosenkranz2025multivariate,
  author  = {Rosenkranz, Matthias and Brunner, Eric and Marin-Sanchez, Gabriel and Fitzpatrick, Nathan and Dilkes, Silas and Tang, Yao and Kikuchi, Yuta and Benedetti, Marcello},
  title   = {Quantum state preparation for multivariate functions},
  journal = {Quantum},
  volume  = {9},
  pages   = {1703},
  year    = {2025}
}

@misc{rattew2022continuous,
  author        = {Rattew, Arthur G. and Koczor, B{\'a}lint},
  title         = {Preparing arbitrary continuous functions in quantum registers with logarithmic complexity},
  year          = {2022},
  eprint        = {2205.00519},
  archivePrefix = {arXiv}
}

@article{low2017qsp,
  author  = {Low, Guang Hao and Chuang, Isaac L.},
  title   = {Optimal {H}amiltonian Simulation by Quantum Signal Processing},
  journal = {Physical Review Letters},
  volume  = {118},
  pages   = {010501},
  year    = {2017}
}

@inproceedings{gilyen2019qsvt,
  author    = {Gily{\'e}n, Andr{\'a}s and Su, Yuan and Low, Guang Hao and Wiebe, Nathan},
  title     = {Quantum singular value transformation and beyond: exponential improvements for quantum matrix arithmetics},
  booktitle = {Proceedings of the 51st Annual ACM SIGACT Symposium on Theory of Computing (STOC)},
  pages     = {193--204},
  year      = {2019}
}

@article{motlagh2024gqsp,
  author  = {Motlagh, Danial and Wiebe, Nathan},
  title   = {Generalized Quantum Signal Processing},
  journal = {PRX Quantum},
  volume  = {5},
  pages   = {020368},
  year    = {2024}
}

@misc{mcardle2022arithmetic,
  author        = {McArdle, Sam and Gily{\'e}n, Andr{\'a}s and Berta, Mario},
  title         = {Quantum state preparation without coherent arithmetic},
  year          = {2022},
  eprint        = {2210.14892},
  archivePrefix = {arXiv}
}

@article{obrien2025piecewise,
  author  = {O'Brien, Oliver and S{\"u}nderhauf, Christoph},
  title   = {Quantum state preparation via piecewise {QSVT}},
  journal = {Quantum},
  volume  = {9},
  pages   = {1786},
  year    = {2025}
}

@article{ran2020mps,
  author  = {Ran, Shi-Ju},
  title   = {Encoding of matrix product states into quantum circuits of one- and two-qubit gates},
  journal = {Physical Review A},
  volume  = {101},
  pages   = {032310},
  year    = {2020}
}

@article{dilip2022compression,
  author  = {Dilip, Rohit and Liu, Yu-Jie and Smith, Adam and Pollmann, Frank},
  title   = {Data compression for quantum machine learning},
  journal = {Physical Review Research},
  volume  = {4},
  pages   = {043007},
  year    = {2022}
}

@misc{jumade2023shortdepth,
  author        = {Jumade, Raghav and Sawaya, Nicolas P. D.},
  title         = {Data is often loadable in short depth: Quantum circuits from tensor networks for finance, images, fluids, and proteins},
  year          = {2023},
  eprint        = {2309.13108},
  archivePrefix = {arXiv}
}

@article{jobst2024mps,
  author  = {Jobst, Bernhard and Shen, Kevin and Riofr{\'i}o, Carlos A. and Shishenina, Elvira and Pollmann, Frank},
  title   = {Efficient {MPS} representations and quantum circuits from the {F}ourier modes of classical image data},
  journal = {Quantum},
  volume  = {8},
  pages   = {1544},
  year    = {2024},
  doi     = {10.22331/q-2024-12-03-1544}
}

@misc{alonsolinaje2025compilation,
  author        = {Alonso-Linaje, Guillermo and Azad, Utkarsh and Soni, Jay and Smalley, Jarrett and Lapworth, Leigh and Arrazola, Juan Miguel},
  title         = {Quantum compilation framework for data loading},
  year          = {2025},
  eprint        = {2512.05183},
  archivePrefix = {arXiv}
}

@misc{zhang2026aqer,
  author        = {Zhang, Kaining and Wang, Xinbiao and Du, Yuxuan and Hsieh, Min-Hsiu and Tao, Dacheng},
  title         = {{AQER}: A Scalable and Efficient Data Loader for Digital Quantum Computers},
  year          = {2026},
  eprint        = {2602.02165},
  archivePrefix = {arXiv}
}

@article{daimon2024transformer,
  author  = {Daimon, Shunsuke and Matsushita, Yu-ichiro},
  title   = {Quantum circuit generation for amplitude encoding using a transformer decoder},
  journal = {Physical Review Applied},
  volume  = {22},
  pages   = {L041001},
  year    = {2024},
  doi     = {10.1103/PhysRevApplied.22.L041001}
}

@misc{aoun2026neural,
  author        = {Aoun, Kevin W. and Kiwit, Florian J. and Riofr{\'i}o, Carlos A. and Saab, Samer, Jr. and Al Bateh, Charbel and Tekli, Joe and Luckow, Andre},
  title         = {Quantum State Preparation via Neural Network Encoding in Quantum Machine Learning},
  year          = {2026},
  eprint        = {2605.31006},
  archivePrefix = {arXiv},
  primaryClass  = {quant-ph}
}

@misc{budiutama2026aiqt,
  author        = {Budiutama, Gekko and Daimon, Shunsuke and Huang, Xinchi and Nishi, Hirofumi and Matsushita, Yu-ichiro},
  title         = {Approximate Amplitude Encoding with the Adaptive Interpolating Quantum Transform},
  year          = {2026},
  eprint        = {2603.03803},
  archivePrefix = {arXiv},
  primaryClass  = {quant-ph}
}

@article{amy2018graysynth,
  author  = {Amy, Matthew and Azimzadeh, Parsiad and Mosca, Michele},
  title   = {On the {CNOT}-complexity of {CNOT}-{PHASE} circuits},
  journal = {Quantum Science and Technology},
  volume  = {4},
  pages   = {015002},
  year    = {2019}
}

@inproceedings{cowtan2020gadgets,
  author    = {Cowtan, Alexander and Dilkes, Silas and Duncan, Ross and Simmons, Will and Sivarajah, Seyon},
  title     = {Phase Gadget Synthesis for Shallow Circuits},
  booktitle = {Proceedings of the 16th International Conference on Quantum Physics and Logic (QPL 2019)},
  series    = {Electronic Proceedings in Theoretical Computer Science},
  volume    = {318},
  pages     = {213--228},
  year      = {2020}
}

@misc{coppersmith1994aqft,
  author        = {Coppersmith, Don},
  title         = {An approximate {F}ourier transform useful in quantum factoring},
  year          = {1994},
  note          = {IBM Research Report RC19642},
  eprint        = {quant-ph/0201067},
  archivePrefix = {arXiv}
}

@article{barenco1996aqft,
  author  = {Barenco, Adriano and Ekert, Artur and Suominen, Kalle-Antti and T{\"o}rm{\"a}, P{\"a}ivi},
  title   = {Approximate quantum {F}ourier transform and decoherence},
  journal = {Physical Review A},
  volume  = {54},
  pages   = {139--146},
  year    = {1996}
}

@misc{javadiabhari2024qiskit,
  author        = {Javadi-Abhari, Ali and Treinish, Matthew and Krsulich, Kevin and Wood, Christopher J. and Lishman, Jake and Gacon, Julien and Martiel, Simon and Nation, Paul D. and Bishop, Lev S. and Cross, Andrew W. and others},
  title         = {Quantum computing with {Qiskit}},
  year          = {2024},
  eprint        = {2405.08810},
  archivePrefix = {arXiv}
}

@article{biamonte2017qml,
  author  = {Biamonte, Jacob and Wittek, Peter and Pancotti, Nicola and Rebentrost, Patrick and Wiebe, Nathan and Lloyd, Seth},
  title   = {Quantum machine learning},
  journal = {Nature},
  volume  = {549},
  pages   = {195--202},
  year    = {2017}
}

@article{preskill2018nisq,
  author  = {Preskill, John},
  title   = {Quantum Computing in the {NISQ} era and beyond},
  journal = {Quantum},
  volume  = {2},
  pages   = {79},
  year    = {2018}
}

@article{navarro2025hifda,
  author  = {Navarro, Victor and Barrag{\'a}n, Marta and Nieto, Rub{\'e}n and Hern{\'a}ndez, {\'A}lvaro and others},
  title   = {{HIFDA} -- High-Frequency Electrical Voltage and Current Signals from Household Appliances},
  journal = {Scientific Data},
  volume  = {12},
  pages   = {527},
  year    = {2025},
  doi     = {10.1038/s41597-025-04859-3},
  
}

@article{medico2020plaid,
  author  = {Medico, Roberto and De Baets, Leen and Gao, Jingkun and Giri, Suman and Kara, Emre and Dhaene, Tom and Develder, Chris and Berg{\'e}s, Mario and Deschrijver, Dirk},
  title   = {A voltage and current measurement dataset for plug load appliance identification in households},
  journal = {Scientific Data},
  volume  = {7},
  pages   = {49},
  year    = {2020},
  doi     = {10.1038/s41597-020-0389-7}
}

@inproceedings{kahl2016whited,
  author    = {Kahl, Matthias and Haq, Anwar Ul and Kriechbaumer, Thomas and Jacobsen, Hans-Arno},
  title     = {{WHITED} -- A Worldwide Household and Industry Transient Energy Data Set},
  booktitle = {3rd International Workshop on Non-Intrusive Load Monitoring (NILM)},
  pages     = {1--4},
  year      = {2016}
}

@inproceedings{lessmeier2016paderborn,
  author    = {Lessmeier, Christian and Kimotho, James Kuria and Zimmer, Detmar and Sextro, Walter},
  title     = {Condition Monitoring of Bearing Damage in Electromechanical Drive Systems by Using Motor Current Signals of Electric Motors: A Benchmark Data Set for Data-Driven Classification},
  booktitle = {Proceedings of the European Conference of the Prognostics and Health Management Society},
  volume    = {3},
  year      = {2016}
}

@article{wagner2020ptbxl,
  author  = {Wagner, Patrick and Strodthoff, Nils and Bousseljot, Ralf-Dieter and Kreiseler, Dieter and Lunze, Fatima I. and Samek, Wojciech and Schaeffter, Tobias},
  title   = {{PTB-XL}, a large publicly available electrocardiography dataset},
  journal = {Scientific Data},
  volume  = {7},
  pages   = {154},
  year    = {2020}
}

@article{goldberger2000physionet,
  author  = {Goldberger, Ary L. and Amaral, Luis A. N. and Glass, Leon and Hausdorff, Jeffrey M. and Ivanov, Plamen Ch. and Mark, Roger G. and Mietus, Joseph E. and Moody, George B. and Peng, Chung-Kang and Stanley, H. Eugene},
  title   = {{PhysioBank}, {PhysioToolkit}, and {PhysioNet}: Components of a New Research Resource for Complex Physiologic Signals},
  journal = {Circulation},
  volume  = {101},
  number  = {23},
  pages   = {e215--e220},
  year    = {2000}
}

@article{cameron2008cluster,
  author  = {Cameron, A. Colin and Gelbach, Jonah B. and Miller, Douglas L.},
  title   = {Bootstrap-Based Improvements for Inference with Clustered Errors},
  journal = {The Review of Economics and Statistics},
  volume  = {90},
  number  = {3},
  pages   = {414--427},
  year    = {2008}
}

@article{golub1973varpro,
  author  = {Golub, Gene H. and Pereyra, Victor},
  title   = {The Differentiation of Pseudo-Inverses and Nonlinear Least Squares Problems Whose Variables Separate},
  journal = {SIAM Journal on Numerical Analysis},
  volume  = {10},
  number  = {2},
  pages   = {413--432},
  year    = {1973}
}

@book{jolliffe2002pca,
  author    = {Jolliffe, Ian T.},
  title     = {Principal Component Analysis},
  edition   = {2},
  publisher = {Springer},
  address   = {New York},
  year      = {2002}
}

@article{donoho2006compressed,
  author  = {Donoho, David L.},
  title   = {Compressed sensing},
  journal = {IEEE Transactions on Information Theory},
  volume  = {52},
  number  = {4},
  pages   = {1289--1306},
  year    = {2006}
}

@article{candes2009matrix,
  author  = {Cand{\`e}s, Emmanuel J. and Recht, Benjamin},
  title   = {Exact matrix completion via convex optimization},
  journal = {Foundations of Computational Mathematics},
  volume  = {9},
  number  = {6},
  pages   = {717--772},
  year    = {2009}
}

@misc{faryad2026inputproblem,
  author        = {Faryad, Muhammad},
  title         = {The Input Problem: A Permanent Bottleneck for Quantum Machine Learning},
  year          = {2026},
  eprint        = {2608.08433},
  archivePrefix = {arXiv}
}

@misc{lin2026distilling,
  author        = {Lin, Guang and Zhao, Qibin},
  title         = {Distilling Datasets into Shallow Circuits for Quantum Machine Learning},
  year          = {2026},
  eprint        = {2609.28229},
  archivePrefix = {arXiv}
}

@misc{zhao2026massive,
  author        = {Zhao, Haimeng and Zlokapa, Alexander and Neven, Hartmut and Babbush, Ryan
                   and Preskill, John and McClean, Jarrod R. and Huang, Hsin-Yuan},
  title         = {Exponential quantum advantage in processing massive classical data},
  year          = {2026},
  eprint        = {2604.07639},
  archivePrefix = {arXiv}
}

@misc{sarmina2026qpipe,
  author        = {Garc{\'i}a Sarmina, Brian and Mart{\'i}nez-Guerrero, Emmanuel
                   and De Anda Gil, Janeth and Guo-Hua, Sun and Shi-Hai, Dong},
  title         = {{Q-PIPE}: a practical quantum phase encoding method},
  year          = {2026},
  eprint        = {2604.09869},
  archivePrefix = {arXiv}
}

@misc{balzano2010subspaces,
  author        = {Balzano, Laura and Nowak, Robert and Recht, Benjamin},
  title         = {Online identification and tracking of subspaces from highly incomplete information},
  year          = {2010},
  eprint        = {1006.4046},
  archivePrefix = {arXiv}
}

@article{candes2007incoherence,
  author  = {Cand{\`e}s, Emmanuel J. and Romberg, Justin},
  title   = {Sparsity and incoherence in compressive sampling},
  journal = {Inverse Problems},
  volume  = {23},
  number  = {3},
  year    = {2007},
  doi     = {10.1088/0266-5611/23/3/008}
}

\end{document}